\documentclass[aps,floatfix,prd,nofootinbib,superscriptaddress,reprint,10pt,preprintnumbers,longbibliography]{revtex4-1}
\pdfoutput=1
\usepackage[utf8]{inputenc}
\usepackage[pdftex]{graphicx}
\usepackage{float}
\usepackage{amsmath}
\usepackage{amssymb}
\usepackage{mathtools}
\usepackage{braket}
\usepackage{siunitx}
\usepackage{amsfonts}
\usepackage{dsfont}
\usepackage{array}
\usepackage{bm}
\usepackage{mathrsfs}
\usepackage{pifont}
\usepackage{multirow}
\usepackage{upgreek}
\usepackage[dvipsnames]{xcolor}
\usepackage[normalem]{ulem}
\usepackage[pdftex,
  pdftitle={Finite-Volume Step Scaling at Fixed Truncation in (2+1)D Hamiltonian U(1) Lattice Gauge Theory},
  pdfauthor={},
  bookmarks,
  colorlinks,
  linkcolor=myblue,
  citecolor=mymagenta,
  menucolor=black,
  urlcolor=myblue,
  plainpages=false,
  pdfpagelabels,
  hypertexnames=false]{hyperref}
\usepackage{orcidlink}
\usepackage{verbatim}
\usepackage{slashed}
\usepackage{cleveref}
\usepackage{subcaption}
\usepackage{csquotes}
\usepackage{bbold}
\usepackage{float}
\makeatletter
\let\newfloat\newfloat@ltx
\makeatother
\usepackage{algorithm}
\makeatletter
\renewcommand{\ALG@name}{Algorithm }
\makeatother
\usepackage[noend]{algpseudocode}
\usepackage{tikz}
\usepackage{tikz-3dplot}
\usetikzlibrary{arrows}
\usetikzlibrary{arrows.meta}
\usetikzlibrary{decorations.markings}
\usetikzlibrary{calc}
\usetikzlibrary{decorations.fractals}
\usetikzlibrary{patterns,angles,quotes}

\newcommand{\Uone}{\mathrm{U}(1)}

\definecolor{mymagenta}{RGB}{200, 0, 100}
\definecolor{myblue}{RGB}{45, 48, 146}
\definecolor{mydarkgreen}{RGB}{0, 90, 45}

\begin{document}
\raggedbottom
\title{Finite-Volume Step Scaling at Fixed Truncation in (2+1)D Hamiltonian \texorpdfstring{$\Uone$}{U(1)} Lattice Gauge Theory}

\author{Alessio Negro\,\orcidlink{0009-0007-0225-6535}}
\email{alessio.negro@hiskp.uni-bonn.de}
\affiliation{Helmholtz-Institut f\"ur Strahlen- und Kernphysik, University of Bonn, Nussallee 14-16, 53115 Bonn, Germany}
\affiliation{Bethe Center for Theoretical Physics, University of Bonn, Nussallee 12, 53115 Bonn, Germany}

\author{Emil Otis Rosanowski\,\orcidlink{0009-0002-4072-2801}}
\email{rosanowski@hiskp.uni-bonn.de}
\affiliation{Helmholtz-Institut f\"ur Strahlen- und Kernphysik, University of Bonn, Nussallee 14-16, 53115 Bonn, Germany}
\affiliation{Bethe Center for Theoretical Physics, University of Bonn, Nussallee 12, 53115 Bonn, Germany}
\affiliation{Transdisciplinary Research Area ``Building Blocks of Matter and Fundamental Interactions'' (TRA Matter), University of Bonn, 53115 Bonn, Germany}

\author{Lena Funcke\,\orcidlink{0000-0001-5022-9506}}
\affiliation{Helmholtz-Institut f\"ur Strahlen- und Kernphysik, University of Bonn, Nussallee 14-16, 53115 Bonn, Germany}
\affiliation{Bethe Center for Theoretical Physics, University of Bonn, Nussallee 12, 53115 Bonn, Germany}
\affiliation{Transdisciplinary Research Area ``Building Blocks of Matter and Fundamental Interactions'' (TRA Matter), University of Bonn, 53115 Bonn, Germany}

\author{Timo~Jakobs\,\orcidlink{0009-0007-5277-797X}}
\affiliation{Helmholtz-Institut f\"ur Strahlen- und Kernphysik, University of Bonn, Nussallee 14-16, 53115 Bonn, Germany}
\affiliation{Bethe Center for Theoretical Physics, University of Bonn, Nussallee 12, 53115 Bonn, Germany}

\author{Karl Jansen\,\orcidlink{0000-0002-1574-7591}}
\affiliation{Deutsches Elektronen-Synchrotron DESY, Platanenallee 6, 15738 Zeuthen, Germany}
\affiliation{Computation-Based Science and Technology Research Center,
The Cyprus Institute, 20 Kavafi Street, 2121 Nicosia, Cyprus}

\author{Paul Ludwig\,\orcidlink{0009-0004-9101-9766}}
\affiliation{Helmholtz-Institut f\"ur Strahlen- und Kernphysik, University of Bonn, Nussallee 14-16, 53115 Bonn, Germany}
\affiliation{Bethe Center for Theoretical Physics, University of Bonn, Nussallee 12, 53115 Bonn, Germany}

\author{Carsten Urbach\,\orcidlink{0000-0003-1412-7582}}
\affiliation{Helmholtz-Institut f\"ur Strahlen- und Kernphysik, University of Bonn, Nussallee 14-16, 53115 Bonn, Germany}
\affiliation{Bethe Center for Theoretical Physics, University of Bonn, Nussallee 12, 53115 Bonn, Germany}

\date{September 23, 2026}

\begin{abstract}
    We propose a finite-volume scheme to perform controlled continuum extrapolations of the lattice step-scaling function, a key ingredient for determining the running coupling in a Hamiltonian lattice gauge theory in small volumes. As a testbed, we employ a dual Hamiltonian formulation of pure $\Uone$ gauge theory in (2+1) dimensions and an operator basis that remains efficient toward weak coupling. We describe the implementation of static external charges on the spatial lattice and study, using matrix product states, the resulting confining string, from which we extract the static potential and a force-based renormalized coupling. The result is a stable lattice-spacing extrapolation at fixed truncation $\Lambda$ and transverse lattice extent $N_y$; a full compact-$\Uone$ continuum result additionally requires $\Lambda$ convergence and control of the transverse physical extent. The method is readily extendable to other gauge groups and dimensions, providing a pathway toward Hamiltonian step-scaling studies in other theories.
\end{abstract}

\maketitle

\section{Introduction}\label{sec:intro}

The Standard Model of particle physics describes the known elementary particles and three of the four fundamental interactions. Among them, the strong interaction, described by quantum chromodynamics (QCD), binds quarks and gluons into the hadrons that make up most of the visible matter in the universe. A defining feature of QCD is that the effective strength of the interaction, encoded in the strong coupling, depends on the energy scale at which it is probed: it becomes weak at short distances (asymptotic freedom) and strong at long distances, where quarks and gluons remain confined inside hadrons. The low-energy regime is inherently non-perturbative, and here the theory dynamically generates a scale $\Lambda_{\text{QCD}}$ that cannot be accessed in perturbation theory.

Wilson's lattice formulation of gauge theories~\cite{Wilson:1974sk} provides the non-perturbative regularization required to directly access the running coupling and $\Lambda_{\text{QCD}}$~\cite{Patrignani:2016xqp}. Connecting a low-energy reference scale to the perturbative high-energy regime is, however, obstructed by the window problem, where the scale that is being probed must simultaneously fulfill $\mu \ll 1/a$ and $1/L \ll \mu$, with $a$ the lattice spacing and $L$ the linear extent of the lattice. The established strategy to overcome this is the step-scaling approach, introduced by Lüscher, Weisz, and Wolff~\cite{Luscher:1991wu}, where controlled changes of scale are iterated in finite volume, with each step extrapolated to the continuum limit~\cite{Luscher:1985dn,Sommer:1993ce,Luscher:1993gh}.

The standard tool for these computations is Markov-chain Monte Carlo (MCMC) sampling, whose limitations for lattice-regularized field theories are well known~\cite{Troyer:2004ge} and directly affect step-scaling studies. In particular, taking the continuum limit leads to autocorrelation times growing exponentially as the lattice spacing is reduced~\cite{Schaefer:2010hu,Schaefer:2010qh}, a phenomenon known as critical slowing down. An alternative is offered by the Hamiltonian formulation of lattice gauge theories (LGTs), introduced by Kogut and Susskind~\cite{Kogut:1974ag}, in which one uses variational methods that are free from autocorrelation, such as tensor networks (TNs)~\cite{Schollwoeck:2010dmrg,PerezGarcia:2007}.

Reaching the continuum limit in this framework is nonetheless challenging, since the weak-coupling regime relevant for continuum extrapolations is the most demanding for TN simulations. Recent developments in the choice of Hamiltonian representation and operator basis now make it possible to keep the relevant terms local and numerically tractable deep into weak coupling~\cite{Jakobs:2022ugr,Jakobs:2025zcv}. A distinct challenge, however, is posed by finite-volume effects, which make the continuum limit of the step-scaling function particularly delicate on the small volumes accessible to current Hamiltonian and prospective quantum platforms.

In this work, we address this challenge by proposing a finite-volume scheme to compute the continuum limit of the lattice step-scaling function, based on the force definition~\cite{Sommer:1993ce} of the renormalized coupling and building on Refs.~\cite{Clemente:2022cka,Crippa:2024cqr}. We demonstrate it on compact $\Uone$ gauge theory in (2+1) dimensions, using TNs in a dual Hamiltonian formulation with static external charges.

With this setup we observe the formation of a confining flux tube between the static charges, from which we extract the static potential and a force-based renormalized coupling. Using the proposed finite-volume scheme, we obtain a stable lattice-spacing extrapolation along a sequence with fixed $r_1/N_x$ and $N_y=3$; only the longitudinal ratio is held fixed, while the transverse physical extent $aN_y$ shrinks. This is not a validated full compact-$\Uone$ continuum result, which would require $\Lambda$ convergence and control of the transverse physical extent.

The paper is organized as follows. Section~\ref{sec:theory} presents the Hamiltonian formulation of LGTs, the step-scaling strategy, and the finite-volume extrapolation strategy. Section~\ref{sec:methods} describes the operator representation and numerical implementation, as well as the TN formulation. Section~\ref{sec:results} presents the numerical results, from the formation of the confining string to the continuum extrapolation of the step-scaling function and the analysis of the leading systematic uncertainties. Section~\ref{sec:discussion} summarizes the implications, and Section~\ref{sec:conclusion} presents the conclusions and outlook. Appendix~\ref{sec:app} contains a detailed transfer-matrix derivation of the dual Hamiltonian for compact $\Uone$ in (2+1) dimensions with static charges. Appendix~\ref{app:uncontrolled} presents a negative-control extrapolation showing that, without the fixed-ratio finite-volume prescription, finite-volume effects dominate the naive continuum limit of the step-scaling function. Appendix~\ref{app:path_deformation} studies finite-cutoff path dependence, Appendix~\ref{app:potential_fit_range} tests the potential-fit range, and Appendix~\ref{app:weak_lambda} examines weak-coupling behavior and local-basis dependence.

\section{Theory}\label{sec:theory}

This section introduces compact $\Uone$ lattice gauge theory in (2+1) dimensions in the Hamiltonian formalism. The continuum theory is regularized on a rectangular spatial lattice, with the time direction kept continuous~\cite{Kogut:1974ag}. Throughout, $N_x\times N_y$ denotes the number of sites of the original lattice, and $a$ its lattice spacing. We label original-lattice sites by the zero-based integer coordinates $(x,y)$, with $x=0,\ldots,N_x-1$ and $y=0,\ldots,N_y-1$. The plaquette whose lower-left original-lattice site has coordinates $(\tilde{x},\tilde{y})$ is represented by the dual-lattice site $\tilde{\mathbf{x}}$; its coordinates run over $\tilde{x}=0,\ldots,N_x-2$ and $\tilde{y}=0,\ldots,N_y-2$. The presentation includes both the definition of the Hamiltonian theory (operators, Hamiltonians, and static charges) and the methodological framework for the step-scaling and finite-volume continuum extrapolation.

\subsection{Operators}

We denote by \(U_{\mathbf{x},k}=e^{i\phi_{\mathbf{x},k}}\in\Uone\) the classical gauge variable on the oriented link from \(\mathbf{x}\) in direction \(k\), and by \(L_{\mathbf{x},k}\) its conjugate momentum.

Upon canonical quantization, the classical variables are promoted to operators \(\hat U_{\mathbf{x},k}\) and \(\hat L_{\mathbf{x},k}\). The Hilbert space of a single link is
\begin{equation}
    \mathcal{H}_{\text{link}}=L^2(\Uone,dU)\,,
\end{equation}
where \(dU\) is the Haar measure on \(\Uone\), and the total Hilbert space is the tensor product over all links
\begin{equation}
    \mathcal{H}_{\text{tot}}=\bigotimes_{\text{links}}\mathcal{H}_{\text{link}}\,.
\end{equation}

A quantum state is therefore represented by a wave function
\begin{equation}
    \psi(\{U_{\mathbf{x},k}\}):\underbrace{\Uone\times\ldots\times\Uone}_{N_{\text{links}}}\to\mathbb{C}\,.
\end{equation}

The operators satisfy the canonical commutation relations
\begin{align}
    [\hat{L}_{\mathbf{x},i},\hat{U}_{\mathbf{y},j}] &= \delta_{\mathbf{x},\mathbf{y}}\,\delta_{i,j}\,\hat{U}_{\mathbf{x},i}\,, \\
    [\hat{L}_{\mathbf{x},i},\hat{U}^{\dagger}_{\mathbf{y},j}] &= -\,\delta_{\mathbf{x},\mathbf{y}}\,\delta_{i,j}\,\hat{U}^{\dagger}_{\mathbf{x},i}\,, \\
    [\hat{L}_{\mathbf{x},i},\hat{L}_{\mathbf{y},j}] &= 0\,.
\end{align}

In the group-element representation, the link operator acts by multiplication
\begin{equation}
    (\hat{U}_{\mathbf{x},k}\psi)(\{U_{\mathbf{x},k}\})
    =U_{\mathbf{x},k}\,\psi(\{U_{\mathbf{x},k}\})
    =e^{i\phi_{\mathbf{x},k}}\psi(\{U_{\mathbf{x},k}\})\,,
\end{equation}
while the conjugate momentum acts as the Lie derivative
\begin{equation}
    \hat{L}_{\mathbf{x},k}=-i\frac{\partial}{\partial\phi_{\mathbf{x},k}}\,.
\end{equation}

The elementary gauge-invariant plaquette operator at site $\mathbf{x}$ in the $(i,j)$-plane reads
\begin{equation}
    \hat{P}_{\mathbf{x},ij}
    =\hat{U}_{\mathbf{x},i}\,
      \hat{U}_{\mathbf{x}+a\mathbf{i},j}\,
      \hat{U}^{\dagger}_{\mathbf{x}+a\mathbf{j},i}\,
      \hat{U}^{\dagger}_{\mathbf{x},j}\,,
    \qquad (i<j)\,.
\end{equation}

\subsection{Hamiltonians}\label{sec:theory_hamiltonians}

\begin{figure}[!htbp]
    \centering
    \resizebox{\linewidth}{!}{%
    \begin{tikzpicture}[scale=1,transform shape, >=stealth]
\path[use as bounding box] (-0.55,-0.55) rectangle (7.9,2.3);

    \draw[->, thick] (-0.35,-0.35) -- (0.45,-0.35) node[below] {$x$};
    \draw[->, thick] (-0.35,-0.35) -- (-0.35,0.45) node[left] {$y$};

    \foreach \y in {0,...,2} {
        \foreach \x in {0,1} {
        \draw[->, thick] (\x+0.05,\y) -- (\x+0.95,\y);
        }
    }

    \foreach \x in {0,1,2} {
        \foreach \y in {0,1} {
        \draw[->, thick] (\x,\y+0.05) -- (\x,\y+0.95);
        }
    }

    \draw[->, thick] (2.2,1) -- (2.6,1);
    
        \foreach \y in {0} {
        \foreach \x in {0,1} {
        \draw[->, thick] (\x+2.8+0.05,\y) -- (\x+2.8+0.95,\y);
        }
    }

    \foreach \x in {0,1,2} {
        \foreach \y in {0,1} {
        \draw[->, thick] (\x+2.8,\y+0.05) -- (\x+2.8,\y+0.95);
        }
    }

        \foreach \x in {0,1}{
        \foreach \y in {0,1}{
            \draw[->, thick, rounded corners=9pt]
            (\x+2.9,\y+0.1) -- (\x+2.9,\y+0.9) -- (\x+2.9+0.8,\y+0.9) -- (\x+2.9+0.8,\y+0.1) -- (\x+2.95,\y+0.1);
            }
        }

        \draw[->, thick] (5.0,1) -- (5.4,1);

        \foreach \x in {0,1}{
            \foreach \y in {0,1}{
            \draw[->, thick, rounded corners=9pt]
            (\x+2.9+2.7,\y+0.1) -- (\x+2.9+2.7,\y+0.9) -- (\x+2.9+2.7+0.8,\y+0.9) -- (\x+2.9+2.7+0.8,\y+0.1) -- (\x+2.95+2.7,\y+0.1);
            }
        }

    \end{tikzpicture}%
    }
    \caption{Dualization of the lattice in the absence of static charges. Link variables of the original lattice (left) are reparametrized in terms of plaquette variables and auxiliary helper links (center).  In the center panel, all vertical links and the lower-boundary horizontal links are the chosen helper-link set, which forms a spanning tree. Gauss' law fixes their electric fluxes in terms of plaquette fluxes, while their pure gauge group variables may be fixed by maximal-tree gauge, leaving one rotor per plaquette (right).}
    \label{fig:dualization_noSC}
\end{figure}

Following Kogut and Susskind~\cite{Kogut:1974ag}, the Hamiltonian in lattice units ($a=1$) reads
\begin{equation}\label{eq:link_hamiltonian}
    \hat{H}_{\text{link}} = \underbrace{\frac{g^2}{2}\sum_{\mathbf{x},k}\left(\hat{L}_{\mathbf{x},k}\right)^{2}}_{\hat{H}_{\text{link},E}}+\underbrace{\frac{1}{g^{2}}\sum_{\mathbf{x},i<j}\left(\mathbb{1}-\operatorname{Re}\hat{P}_{\mathbf{x},ij}\right)}_{\hat{H}_{\text{link},M}}\,.
\end{equation}

Here, $\operatorname{Re}\hat P=(\hat P+\hat P^\dagger)/2$. Thus the magnetic term in Eq.~\eqref{eq:link_hamiltonian} is equivalently $(2g^2)^{-1}\sum_{\mathbf{x},i<j}(2-\hat P_{\mathbf{x},ij}-\hat P^\dagger_{\mathbf{x},ij})$.

We distinguish the dimensionful continuum coupling $g_{\mathrm{cont}}^2$ from the dimensionless bare lattice coupling $g_{\mathrm{lat}}^2$. In $(2+1)$ dimensions, $[g_{\mathrm{cont}}^2]=1$, and, at tree level, restoring the lattice spacing in Eq.~\eqref{eq:link_hamiltonian} gives $g_{\mathrm{lat}}^2=a g_{\mathrm{cont}}^2$. Throughout the remainder of this work, the symbol $g^2$ denotes $g_{\mathrm{lat}}^2$. The renormalized force-scheme coupling $\overline{g}^2$ is dimensionless.

The link Hamiltonian splits into an electric term $\hat{H}_{\text{link},E}$ and a magnetic term $\hat{H}_{\text{link},M}$; the latter is a four-link interaction in this link formulation.

The total Hilbert space $\mathcal{H}_{\text{tot}}$ is reduced to  $\mathcal{H}_{\text{phys}} \subset \mathcal{H}_{\text{tot}}$ by Gauss' law, which enforces local conservation of electric flux and removes unphysical gauge-variant states~\cite{Bauer:2023jvw}. It reads
\begin{equation}\label{eq:gausslaw}
    \begin{aligned}
      &\hat{G}_{\mathbf{x}}\equiv \sum_{k}\left(\hat{L}_{\mathbf{x},k} - \hat{L}_{\mathbf{x} - a\mathbf{k},k}\right)\,, \\
      & \quad \hat{G}_{\mathbf{x}}\ket{\psi}=0\;\forall\mathbf{x}\,,\quad \ket{\psi }\in \mathcal{H}_{\text{phys}}\,.
    \end{aligned}
\end{equation}

Since $\hat{H}_{\text{link},M}$ dominates as $g^2 \to 0$, the four-link interaction generates substantial ground-state entanglement at weak coupling. To mitigate this, we perform a duality transformation of the Kogut--Susskind Hamiltonian, obtaining a plaquette formulation in which the magnetic term becomes local~\cite{Unmuth-Yockey:2018xak,Weber:2013bea}.

The dual Hamiltonian reads
\begin{equation}\label{eq:dual_hamiltonian}
    \begin{aligned}
    \hat{H}_{\text{dual}}={}&\underbrace{\frac{g^{2}}{2}\sum_{\mathbf{x},i}\left(\hat{L}_{\mathbf{x}}-\hat{L}_{\mathbf{x}+a\mathbf{i}}\right)^{2}}_{\hat{H}_{\text{dual},E}}\\
    &+\underbrace{\frac{1}{2g^2}\sum_{\mathbf{x}}\left(2-\hat{P}_{\mathbf{x}}-\hat{P}^{\dagger}_{\mathbf{x}}\right)}_{\hat{H}_{\text{dual},M}}\,.
    \end{aligned}
\end{equation}

Now the first term is a nearest-neighbor interaction between integer-valued electric fluxes~\cite{Banks:1977cc}, whereas the second term is now local: $\hat{P}_{\mathbf{x}}$ is a plaquette operator residing on the dual site $\mathbf{x}$.

We use open (closed-flux) boundary conditions in both spatial directions, so no non-contractible loops wrap the lattice. The dualization is illustrated in Fig.~\ref{fig:dualization_noSC}: each elementary loop's four links are reparametrized into a plaquette variable plus auxiliary helper links. Together with one plaquette holonomy per face, the tree links give an invertible parametrization of the original link variables, with the interior horizontal links reconstructed recursively. Once the plaquette fluxes are specified, Gauss' law at a tree leaf determines the electric flux on its single incident tree link; removing the leaf and repeating determines all tree-link fluxes. The counting of the number of degrees of freedom is consistent: for an open $N_x\times N_y$ lattice
\begin{equation}
    N_{\text{links}}-N_{\text{sites}}+1=(N_x-1)(N_y-1)=N_{\text{plaq}}\,,
\end{equation}
and the flux on each link is recovered as the oriented difference of the two adjacent plaquette variables.

\subsection{Static charges}\label{sec:theory_electriccharges}

\begin{figure}[!htbp]
    \centering
    \resizebox{\linewidth}{!}{%
    \begin{tikzpicture}[scale=1, transform shape, >=stealth]
\path[use as bounding box] (-0.55,-0.55) rectangle (7.9,2.3);
    \draw[->, thick] (-0.35,-0.35) -- (0.45,-0.35) node[below] {$x$};
    \draw[->, thick] (-0.35,-0.35) -- (-0.35,0.45) node[left] {$y$};

    \filldraw [red] (0,1) circle (3pt);
    \filldraw [blue] (2,1) circle (3pt);

    \draw[->, thick, dashed] (0.05,1) -- (0.95,1);
    \draw[->, thick, dashed] (1.05,1) -- (1.95,1);

    \draw[->, thick, dashed] (0.05 + 2.8,1) -- (0.95 + 2.8,1);
    \draw[->, thick, dashed] (1.05 + 2.8,1) -- (1.95 + 2.8,1);

    \draw[->, thick, dashed] (0.05 + 5.6,1) -- (0.95 + 5.6,1);
    \draw[->, thick, dashed] (1.05 + 5.6,1) -- (1.95 + 5.6,1);

    \filldraw [red] (2.8,1) circle (2.5pt);
    \filldraw [blue] (4.8,1) circle (2.5pt);
        
    \filldraw [red] (5.6,1) circle (2.5pt);
    \filldraw [blue] (7.6,1) circle (2.5pt);

    \foreach \y in {0,2} {
        \foreach \x in {0,1} {
        \draw[->, thick] (\x+0.05,\y) -- (\x+0.95,\y);
        }
    }

    \foreach \x in {0,1,2} {
        \foreach \y in {0,1} {
        \draw[->, thick] (\x,\y+0.05) -- (\x,\y+0.95);
        }
    }

    \draw[->, thick] (2.2,1) -- (2.6,1);
    
        \foreach \y in {0} {
        \foreach \x in {0,1} {
        \draw[->, thick] (\x+2.8+0.05,\y) -- (\x+2.8+0.95,\y);
        }
    }

    \foreach \x in {0,1,2} {
        \foreach \y in {0,1} {
        \draw[->, thick] (\x+2.8,\y+0.05) -- (\x+2.8,\y+0.95);
        }
    }

        \foreach \x in {0,1}{
        \foreach \y in {0,1}{
            \draw[->, thick, rounded corners=9pt]
            (\x+2.9,\y+0.1) -- (\x+2.9,\y+0.9) -- (\x+2.9+0.8,\y+0.9) -- (\x+2.9+0.8,\y+0.1) -- (\x+2.95,\y+0.1);
            }
        }

        \draw[->, thick] (5.0,1) -- (5.4,1);

        \foreach \x in {0,1}{
            \foreach \y in {0,1}{
            \draw[->, thick, rounded corners=9pt]
            (\x+2.9+2.7,\y+0.1) -- (\x+2.9+2.7,\y+0.9) -- (\x+2.9+2.7+0.8,\y+0.9) -- (\x+2.9+2.7+0.8,\y+0.1) -- (\x+2.95+2.7,\y+0.1);
            }
        }

    \end{tikzpicture}%
    }
    \caption{Dualization in the presence of static charges. Gauss' law is modified by external sources and can no longer eliminate all helper links; instead, a path $\gamma$ connecting the charges must be introduced to carry the associated flux (dashed line).}
    \label{fig:dualization_SC}
\end{figure}
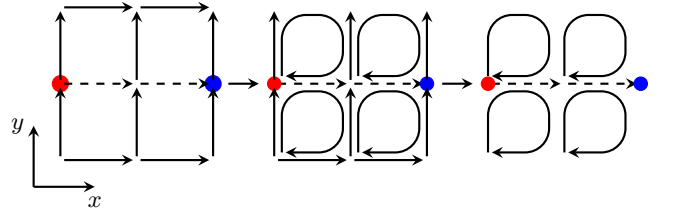

Static charges are non-dynamical sources of electric flux that modify Gauss' law to read
\begin{equation}\label{eq:gausslaw_SC}
    \begin{aligned}
      &\hat{G}_{\mathbf{x}}\equiv \sum_{k}\left(\hat{L}_{\mathbf{x},k} - \hat{L}_{\mathbf{x} - a\mathbf{k},k}\right) + \hat{Q}_\mathbf{x}\,, \\
      & \quad \hat{G}_{\mathbf{x}}\ket{\psi}=0\;\forall\mathbf{x}\,,\quad \ket{\psi }\in \mathcal{H}_{\text{phys}}\,,
    \end{aligned}
\end{equation}
where
\begin{equation}
        \hat{Q}_{\mathbf{x}}=
\begin{cases}
    \pm 1 & \text{positive/negative static charge at $\mathbf{x}$}\,, \\
    0 & \text{no static charge at $\mathbf{x}$}\,. \\
\end{cases}
\end{equation}

With static sources, the helper links cannot be fully gauged out: the charges must be connected by an arbitrary path of helper links $\gamma$ carrying nonzero flux (Fig.~\ref{fig:dualization_SC}). The resulting dual Hamiltonian, derived in Appendix~\ref{sec:app}, reads
\begin{equation}
    \label{eq:dual_hamiltonian_sc}
    \begin{aligned}
        \hat{H}_{\text{dual}}=&\,\frac{g^2}{2}\sum_{\mathbf{x},i} \left(\hat{L}_{\mathbf{x}}-\hat{L}_{\mathbf{x}+a \mathbf{i}} +  n^{(\gamma)}_{\mathbf{x},i}\right)^2+\\
    &\,+\frac{1}{2g^2}\sum_{\mathbf{x}}\left(2-\hat{P}_{\mathbf{x}}-\hat{P}^{\dagger}_{\mathbf{x}}\right)\,.
    \end{aligned}
\end{equation}

For an oriented original-lattice link $\ell$, let $i$ and $j$ be the plaquettes to its left and right; the exterior plaquette field is zero. Write $\hat L_\ell=\hat L_i-\hat L_j+n^{(\gamma)}_{ij}$, with $n^{(\gamma)}_{ij}=-q$ if $\ell$ follows $\gamma$, $+q$ if it opposes $\gamma$, and zero otherwise ($q=1$). Reversing $\ell$ exchanges $i,j$ and gives $n^{(\gamma)}_{ji}=-n^{(\gamma)}_{ij}$.

In all production calculations, $\gamma$ is the straight horizontal path between the equal-$y$ static charges, oriented from the positive to the negative source. Among the paths tested at finite $\Lambda$, it gives the lowest optimized charged-sector energy. Increasing $\Lambda$ lowers these variational upper bounds and removes their differences, as shown in Appendix~\ref{app:path_deformation}; in the untruncated theory all paths are exactly equivalent.

\subsection{The step-scaling procedure}\label{sec:theory_latticestepscaling}

For a renormalized coupling defined from the static force, it is natural to formulate step-scaling in terms of a distance scale $r$. The continuum step-scaling function $\sigma_s$ is defined by
\begin{equation}
\overline{g}^2\left(sr\right)=\sigma_s\!\left(\overline{g}^2\left(r\right)\right)\,.
\end{equation}
It describes how the renormalized coupling $\overline{g}$ changes when the reference distance is increased by a factor \(s\).

At finite lattice spacing $a$, one analogously defines the lattice step-scaling function
\begin{equation}\label{eq:matching}
\Sigma_s\left(u,a/r\right)\equiv \overline{g}^2\left(sr,a\right)\Big|_{\overline{g}^2\left(r,a\right)=u}\,.
\end{equation}
To determine $\sigma_s\left(u\right)$, one considers several lattice resolutions $a/r$, tunes the bare coupling on each lattice so that the coupling at the reference distance $r$ takes the common value $u$, and then measures the coupling at the stepped distance $sr$ at the same tuned bare coupling. The continuum step-scaling function is obtained from
\begin{equation}
\sigma_s\left(u\right)=\lim_{a/r\to 0}\Sigma_s\left(u,a/r\right)\,.
\end{equation}

Iterating $\sigma_s$ connects the coupling at the sequence of scales $r, sr, s^2r,\dots$ In the present work, the practical implementation of this program requires controlling finite-volume effects while keeping the physical reference distance fixed; this is described in the next subsection.

\subsection{Finite-volume continuum extrapolation}\label{sec:theory_FVS}

Taking the continuum limit of $\Sigma_s$ requires sending $a\to 0$ at fixed physical scale, i.e.\ increasing the resolution $r_1/a$ of the reference distance. The most direct way to do this, raising $r_1/a$ without enlarging the lattice in proportion, changes the physical volume (measured in units of $r_1$) from one lattice in the sequence to the next. On the small volumes accessible to Hamiltonian simulations this finite-volume contribution is not negligible, so it must be controlled along the sequence rather than ignored.

We hold the longitudinal ratio $r_1/N_x$ fixed along the continuum sequence, preventing changes in the longitudinal geometry from being mixed with cutoff dependence. Otherwise, the measured $\Sigma_s$ picks up a varying finite-volume contribution on top of the cutoff dependence, producing an irregular sequence rather than a smooth extrapolation (see Appendix~\ref{app:uncontrolled}).

We define the dimensionless renormalized force-scheme coupling by
\begin{equation}
\overline{g}^2\left(r\right) = r^2 \frac{dV\left(r\right)}{dr} = r^2 F\left(r\right)\,.
\end{equation}

We adopt this force-based scheme because it is naturally accessible in the Hamiltonian formulation, as the vacuum and charged-sector ground-state energies are obtained directly, so the static potential does not require extracting the large-time decay of Euclidean correlation functions. Furthermore, fitting the potential allows the force to be evaluated at continuously variable separations, which is required to keep $r_1/N_x$ fixed along the continuum sequence.

For a given lattice and bare coupling, we first compute the static potential from the vacuum-subtracted ground-state energy of a charge-anticharge pair at separation $r/a$
\begin{equation}\label{eq:pot_from_energy}
aV\left(r/a\right) = a\left(E_{Q\bar Q}\left(r/a\right) - E_0\right)\,.
\end{equation}
We then fit the potential with the ansatz
\begin{equation}\label{eq:potential_fit}
aV\left(r/a\right) = \alpha + \sigma\, r/a + \delta \log\left(r/a\right)\,.
\end{equation}
This ansatz captures the linear confinement at large $r$ and the logarithmic Coulomb behavior expected in (2+1)D at short distances, following standard static-potential fits in compact $\Uone$ gauge theory~\cite{Loan:2002ej,Gross:2025qae}. Differentiating analytically yields the force in lattice units
\begin{equation}\label{eq:force_fit}
a^2 F\left(r/a\right) = \sigma + \frac{\delta}{r/a}\,.
\end{equation}

We now choose a starting lattice with volume $V^{(1)}$, a reference distance $r_1^{(1)}$, and a stepped distance
\begin{equation}
r_2^{(1)} = s\, r_1^{(1)}\,.
\end{equation}
For a sequence of larger lattices $V^{(i)}$, we define
\begin{equation}
r_1^{(i)} = r_1^{(1)} \frac{V^{(i)}}{V^{(1)}}\,,
\qquad
r_2^{(i)} = s\, r_1^{(i)}\,,
\end{equation}
so that
\begin{equation}
\frac{r_1^{(i)}}{V^{(i)}} \equiv \mathrm{const.}\,,
\end{equation}
throughout the sequence. Thus, as the resolution is increased, the distance in lattice units grows while the corresponding physical ratio is kept fixed.

For each lattice $i$, we then tune the bare coupling $g_{(i)}^2$ so that the renormalized coupling at the reference distance is the same on all lattices
\begin{equation}
r_1^{(i)\,2} F\!\left(r_1^{(i)}, g_{(i)}^2\right)
=
r_1^{(j)\,2} F\!\left(r_1^{(j)}, g_{(j)}^2\right)\,,
\qquad
\forall\, i,j\,.
\end{equation}
This enforces a common reference coupling, i.e., the same physical scale across different discretizations.

In our setup, the longitudinal ratio $r_1/N_x$ is held fixed while $N_y=3$ throughout; this controls the longitudinal finite-volume geometry, but the transverse physical size $aN_y$ shrinks as $a\to0$, so finite-$N_y$ effects remain in the continuum sequence and are assessed separately in Section~\ref{sec:results}.
Once this tuning has been performed, the lattice step-scaling observable is
\begin{equation}
\Sigma_s^{(i)} = r_2^{(i)\,2} F\!\left(r_2^{(i)}, g_{(i)}^2\right)\,,
\end{equation}
 {and we extrapolate $\Sigma_s^{(i)}$ in $(a/r_1)^2$ at fixed $\Lambda$ and $N_y$.}

To interpret the scale dependence of this observable, consider first
the tree-level infinite-volume limit. In two spatial dimensions the Coulomb
potential is logarithmic~\cite{Loan:2002ej},
\begin{equation}
V_{\mathrm{tree}}(r)
=V_0+\frac{g_{\mathrm{cont}}^2}{2\pi}
\log\left(\frac{r}{r_0}\right),
\end{equation}
and therefore
\begin{equation}
F_{\mathrm{tree}}(r)=\frac{g_{\mathrm{cont}}^2}{2\pi r},
\qquad
\overline{g}_{\mathrm{tree}}^2(r)
=\frac{g_{\mathrm{cont}}^2r}{2\pi}.
\end{equation}
Thus
$\overline{g}_{\mathrm{tree}}^2(sr)
=s\,\overline{g}_{\mathrm{tree}}^2(r)$ and
$\sigma_s^{\mathrm{tree}}(u)=su$.

For the fitted force, writing $x\equiv r/a$ gives
\begin{equation}
\overline{g}^2(r)
=\sigma x^2+\delta x\,.
\end{equation}
Consequently,
\begin{equation}
u=\sigma x^2+\delta x,
\qquad
\Sigma_s(u,a/r)=s^2\sigma x^2+s\delta x,
\end{equation}
so that
\begin{equation}
\Sigma_s(u,a/r)-su=s(s-1)\sigma x^2\,.
\end{equation}
The logarithmic term retains the tree-level factor $s$, whereas the fitted
linear term scales as $s^2$. Finite volume and open boundaries modify the
infinite-volume baseline itself, while finite $\Lambda$ can further alter the
extracted potential through the Hilbert-space projection. A deviation from
$su$ therefore does not by itself establish a quantum or nonperturbative
contribution.

\section{Numerical methods}\label{sec:methods}

Our numerical methods follow the ones used in Ref.~\cite{Jakobs:2025zcv} closely; we summarize the salient points here.

\subsection{Operator representation and numerical implementation}\label{sec:methods_setup}

We represent operators in the plaquette-state basis, the eigenbasis of the single-plaquette dual Hamiltonian, truncated to the lowest $n=2\Lambda+1$ states. Matrix elements of $\hat L$, $\hat L^2$, and $\hat P$ are then evaluated by numerical integration.

Integrations are performed with \texttt{scipy.integrate.quad} at absolute tolerance $\texttt{epsabs}=\mathcal{O}(10^{-8})$. At small $g^2$ the relevant matrix elements become comparable to \texttt{epsabs}, which practically caps our simulations at $g^2\sim\mathcal{O}(10^{-2})$. Smaller $g^2$ would require tightening \texttt{epsabs} at the cost of integration stability. We test local-basis convergence at $g^2=0.01$ in Appendix~\ref{app:weak_lambda}.

With this setup, the production simulations used for the continuum extrapolation reach lattices up to $N_x=12$. We also tested the scalability of the approach at $\Lambda=1$ and $N_y=4$, finding that lattices up to $N_x=24$ can be reached with increased wall-clock time. This is consistent with the column-major MPS mapping used for the tensor-network optimization (Sec.~\ref{sec:methods_TN}), whose cost is much milder in the longitudinal extent $N_x$ than in the transverse width $N_y$: square lattices up to $7\times 7$ are substantially more demanding, with the largest runs requiring up to one week of wall-clock time on a 96-core large-memory node of the Marvin cluster at the University of Bonn. The finite-volume effects from the smaller transverse direction are assessed a posteriori in Sec.~\ref{sec:results_finitesize}.

 The main-text sequence uses $\Lambda=1$ and $N_y=3$. Appendix~\ref{app:weak_lambda} compares it with $\Lambda=2$ sequences at $N_y=3,5$ and extends the result to $N_x=16,20$.

\subsection{Tensor network formulation}\label{sec:methods_TN}

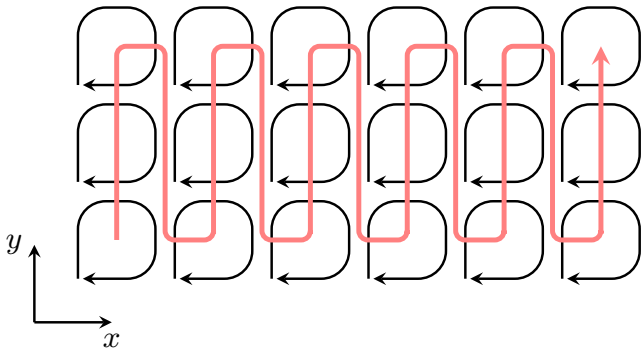
\begin{figure}[!htbp]
    \centering
    \resizebox{\linewidth}{!}{%
    \begin{tikzpicture}[scale=1.1, transform shape, >=stealth]
\path[use as bounding box] (-0.55,-0.55) rectangle (6.1,3.1);

        \draw[->, thick] (-0.35,-0.35) -- (0.45,-0.35) node[below] {$x$};
        \draw[->, thick] (-0.35,-0.35) -- (-0.35,0.45) node[left] {$y$};

        \foreach \x in {0,...,5}{
            \foreach \y in {0,1,2}{

            \draw[->, thick, rounded corners=9pt]
            (\x+0.1,\y+0.1) -- (\x+0.1, \y+0.9) -- (\x+0.9,\y+0.9) -- (\x+0.9,\y+0.1) -- (\x+0.15,\y+0.1);

            }
        }

        \draw[->, red!50, ultra thick, rounded corners=3pt]
        (0.5,0.5) -- (0.5,1.5) -- (0.5,2.5) -- (1,2.5) -- (1,0.5) -- (1.5,0.5) --
        (1.5,0.5) -- (1.5,1.5) -- (1.5,2.5) -- (2,2.5) -- (2,0.5) -- (2.5,0.5) --
        (2.5,0.5) -- (2.5,1.5) -- (2.5,2.5) -- (3,2.5) -- (3,0.5) -- (3.5,0.5) --
        (3.5,0.5) -- (3.5,1.5) -- (3.5,2.5) -- (4,2.5) -- (4,0.5) -- (4.5,0.5) --
        (4.5,0.5) -- (4.5,1.5) -- (4.5,2.5) -- (5,2.5) -- (5,0.5) -- (5.5,0.5) --
        (5.5,0.5) -- (5.5,1.5) -- (5.5,2.5);

    \end{tikzpicture}%
    }
    \caption{The two-dimensional plaquette degrees of freedom are mapped to a one-dimensional MPS chain in column-major order.}
  \label{fig:snakeMPS}
\end{figure}

We determine ground-state properties with a matrix-product-state (MPS) ansatz optimized by two-site DMRG, implemented in the \texttt{ITensor} library~\cite{Fishman:2022itensor,Fishman:2022itensor_release,Stoudenmire:2012dmrg2d}.

We map the two-dimensional dual lattice to a one-dimensional MPS by arranging its plaquette degrees of freedom in the column-major order shown in Fig.~\ref{fig:snakeMPS}. The dual-lattice site $\tilde{\mathbf{x}}$ with coordinate pair $(\tilde{x},\tilde{y})$ is assigned the one-based MPS site index
\begin{equation}
    i_{\text{MPS}}=\tilde{x}(N_y-1)+\tilde{y}+1.
\end{equation}
Here $\tilde{x}=0,\ldots,N_x-2$ and $\tilde{y}=0,\ldots,N_y-2$ are the zero-based dual-lattice coordinates defined in Sec.~\ref{sec:theory}, so $i_{\text{MPS}}=1,\ldots,(N_x-1)(N_y-1)$.

This 2D-to-1D mapping makes the computational cost scale mildly with $N_{x}$ but exponentially with the transverse width $N_{y}$, which limits accessible lattice geometries. In \texttt{ITensor}, singular values below a parameter \texttt{cutoff} are discarded; the remaining count defines the bond dimension $\chi$. We push the SVD cutoff as low as $10^{-15}$ to resolve the small discretization effects relevant for the continuum extrapolation.

We employ a staged optimization strategy for DMRG. We begin with a coarse preconditioning phase consisting of at least $\mathcal{O}(10)$ two-site DMRG sweeps (i.e. full optimization passes in both directions) using a relatively loose cutoff of $10^{-5}$, while gradually increasing the bond-dimension cap and injecting noise to avoid metastable minima. This is followed by an intermediate high-precision run, in which the cutoff is tightened to $10^{-14}$. Finally, a convergence run is performed with a cutoff of $10^{-15}$ and a bond-dimension cap of $\chi_{\max} \sim 500\,N_x N_y (2\Lambda + 1)$. In the simulations reported here, this bond-dimension cap is not saturated in practice. For the system sizes considered, convergence tests show that the ground-state energy $E_0$ is stable at a level comparable to the final SVD cutoff.

As a validation, we independently implement exact diagonalization for the standard Kogut--Susskind link Hamiltonian in the electric-flux basis. For small lattices, Gauss' law can be imposed explicitly (at the cost of nonlocal couplings), which is acceptable for sparse diagonalization. We benchmark energies and selected observables for lattices up to $5\times 5$, finding agreement with the tensor-network results at the level of machine precision.

\section{Results}\label{sec:results}

We now present our numerical results, from the formation of the confining string and the tuning of the bare coupling to the continuum extrapolation of the step-scaling function and the analysis of the leading systematic uncertainties. Their implications and the role of the different systematic effects are discussed further in Sec.~\ref{sec:discussion}.

\subsection{Formation of a confining string}\label{sec:results_string}

\begin{figure}[!htbp]
\centering
    \begin{tikzpicture}
        \node[inner sep=0] (fluxfig) {\includegraphics[width=0.97\linewidth]{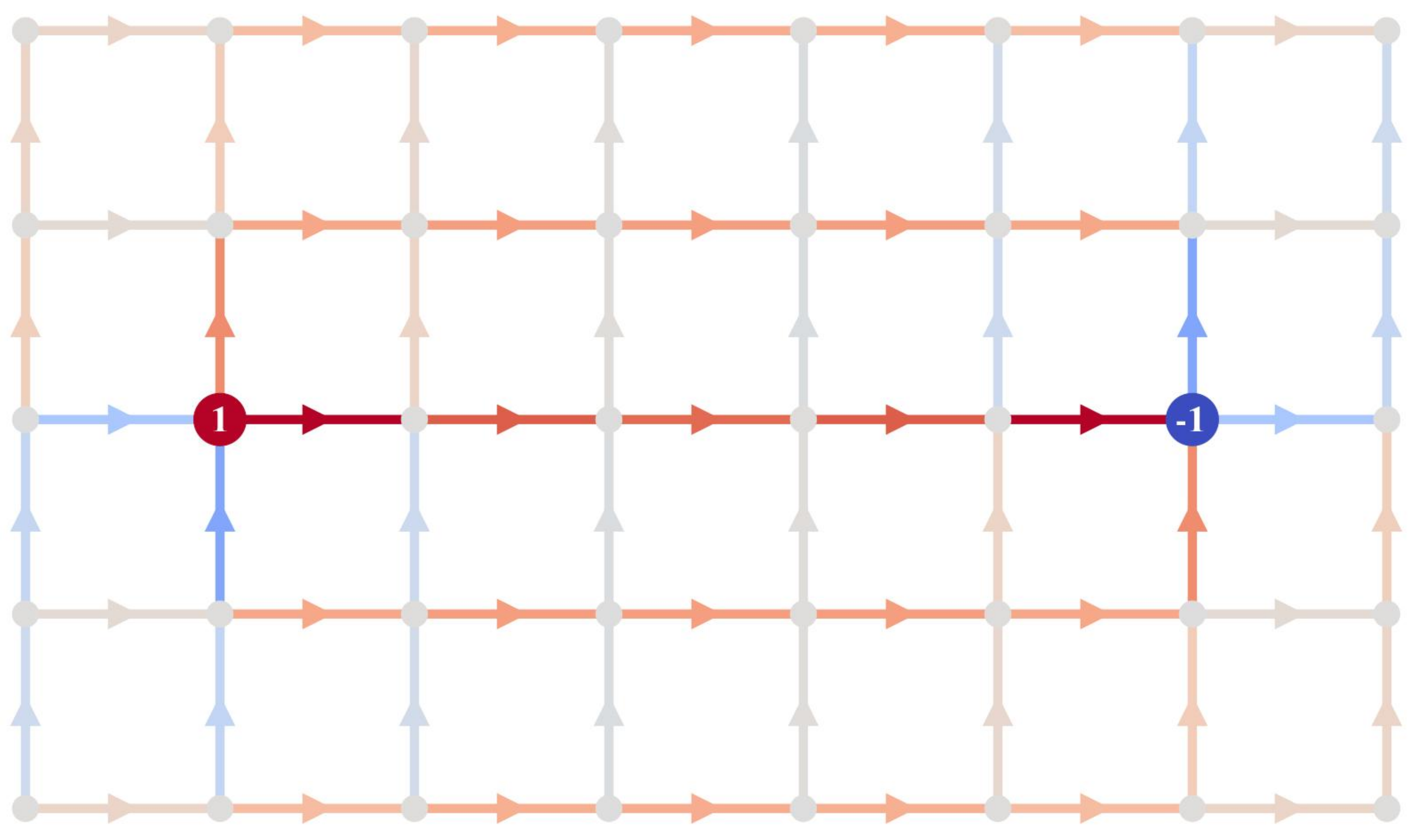}};
        \begin{scope}[shift={($(fluxfig.south west)+(0.18,0.18)$)}]
            \draw[->, thick] (0,0) -- (0.55,0) node[below] {$x$};
            \draw[->, thick] (0,0) -- (0,0.55) node[left] {$y$};
        \end{scope}
    \end{tikzpicture}
    \caption{Formation of a charged flux tube between two static external sources of opposite sign (red: +1, blue: -1). The electric-energy density localizes along the string connecting the charges.}
    \label{fig:string}
\end{figure}

We probe the confining string by placing two opposite charges at original-lattice sites $(x_1,y_0)$ and $(x_2,y_0)$ on a horizontal path, where $y_0=\lfloor (N_y-1)/2\rfloor$ and $1\le x_1<x_2\le N_x-2$. Thus both charges remain at least one lattice spacing from the longitudinal boundaries, and their separation is $r/a=x_2-x_1$, with $1\le r/a\le N_x-3$. Only horizontal axial strings are used in the numerical results.

The electric-energy contribution of a link $l$ adjacent to plaquettes $i$ and $j$ reads
\begin{equation}
    \epsilon_{l}\propto \left\langle \left(\hat L_{i}-\hat L_{j}+n^{(\gamma)}_{l}\right)^{2}\right\rangle\,,
\end{equation}
where $i,j$ label the dual sites of the two plaquettes adjacent to link $l$, and $n^{(\gamma)}_{l}$ is the background-field value on that link. The spatial profile of $\epsilon_l$ is plotted in Fig.~\ref{fig:string} on an $8\times 5$ lattice at $g^2=0.5$, for two charges separated by an axial distance $r=5$. The electric flux is localized along the path joining the sources and decays away from it, as expected for a confining string.

\begin{figure}[!htbp]
\centering
    \includegraphics[width=\linewidth]{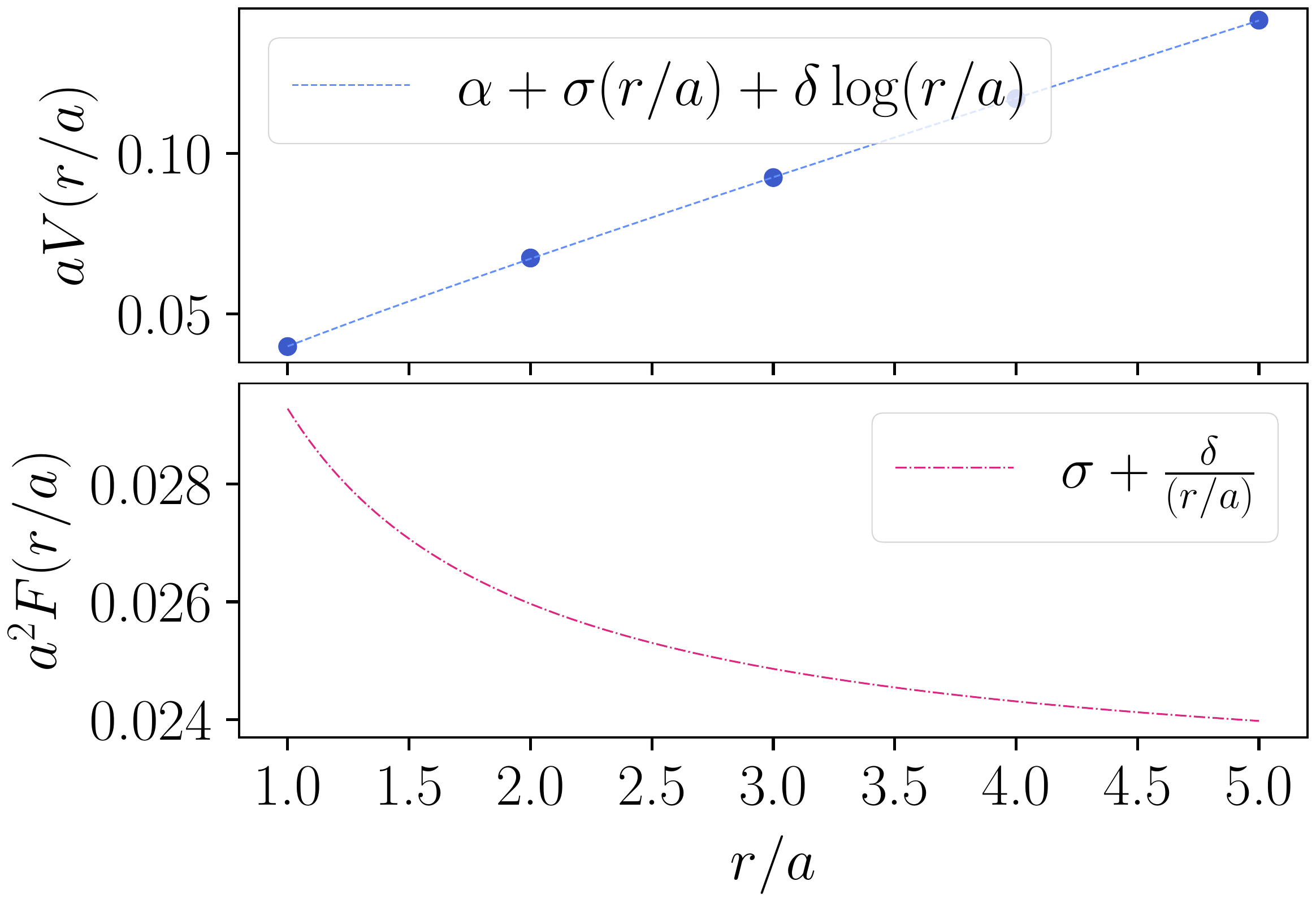}
    \caption{Potential (top) and force (bottom) between two static charges as a function of the separation $r/a$ on an  $8\times 3$ lattice at $g^2=0.14$ and $\Lambda=1$.}
    \label{fig:combined}
\end{figure}

Having established string formation, we extract the vacuum-subtracted ground-state energy of the static charge-anticharge pair at each separation $r$, yielding the intercharge potential $aV\left(r/a\right)$ via Eq.~\eqref{eq:pot_from_energy}.

The production potential fit includes $r/a=1,\ldots,5$.

The fitted potential and derived force are shown in Fig.~\ref{fig:combined}. The fit yields  $\alpha=0.01731$, $\sigma=0.02266$, and $\delta=0.006618$; over the range $r/a=2,\ldots,5$, the logarithmic Coulomb contribution $\delta \log\left(r/a\right)$ amounts to about  $9$--$11\%$ of the linear term $\sigma\,r/a$, and is therefore only weakly visible.

\subsection{Tuning of the bare coupling \texorpdfstring{$g^2$}{g2}}\label{sec:results_tuning}

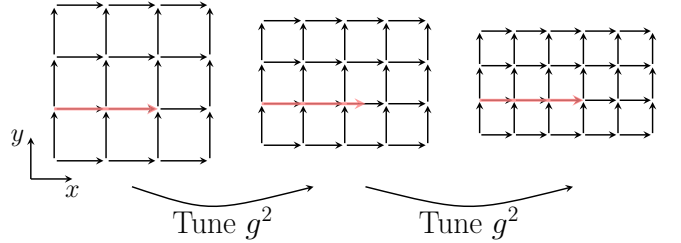
\begin{figure}[!htbp]
    \centering
    \resizebox{\linewidth}{!}{%
    \begin{tikzpicture}[
    transform shape,
    >=stealth,
    lattice/.style={->, black, thick},
    flux/.style={->, ultra thick, red!50, opacity=0.8},
    tunelabel/.style={font=\LARGE}
]

\draw[->, thick] (-0.45,-0.35) -- (0.35,-0.35) node[below, font=\Large] {$x$};
\draw[->, thick] (-0.45,-0.35) -- (-0.45,0.45) node[left, font=\Large] {$y$};

\begin{scope}[shift={(0,0)}, scale=1]
    \foreach \y in {0,1,2,3} {
        \foreach \x in {0,1,2} {
            \draw[lattice] (\x+0.05,\y) -- (\x+0.95,\y);
        }
    }

    \foreach \x in {0,1,2,3} {
        \foreach \y in {0,1,2} {
            \draw[lattice] (\x,\y+0.05) -- (\x,\y+0.95);
        }
    }

    \draw[flux] (0,1) -- (2,1);
\end{scope}

\begin{scope}[shift={(4,0.3)}, scale=0.8]
    \foreach \y in {0,1,2,3} {
        \foreach \x in {0,1,2,3} {
            \draw[lattice] (\x+0.05,\y) -- (\x+0.95,\y);
        }
    }

    \foreach \x in {0,1,2,3,4} {
        \foreach \y in {0,1,2} {
            \draw[lattice] (\x,\y+0.05) -- (\x,\y+0.95);
        }
    }

    \draw[flux] (0,1) -- (2.5,1);
\end{scope}

\begin{scope}[shift={(8.2,0.5)}, scale=2/3]
    \foreach \y in {0,1,2,3} {
        \foreach \x in {0,1,2,3,4} {
            \draw[lattice] (\x+0.05,\y) -- (\x+0.95,\y);
        }
    }

    \foreach \x in {0,1,2,3,4,5} {
        \foreach \y in {0,1,2} {
            \draw[lattice] (\x,\y+0.05) -- (\x,\y+0.95);
        }
    }

    \draw[flux] (0,1) -- (3,1);
\end{scope}

\draw[->, thick]
    (1.5,-0.5) .. controls (3,-1) .. (5,-0.5);
\node[tunelabel] at (3.25,-1.2) {Tune $g^2$};

\draw[->, thick]
    (6,-0.5) .. controls (8,-1) .. (10,-0.5);
\node[tunelabel] at (8,-1.2) {Tune $g^2$};

\end{tikzpicture}%
    }
    \caption{Schematic of the finite-volume continuum sequence. The fixed ratio, chosen here for illustration, is $r_1^{(i)}/N_x^{(i)}=1/2$, with the physical reference distance kept fixed while $N_x$ is increased at fixed $N_y=3$; the values $r_1^{(i)}$ quoted below are in lattice units and grow as $a$ shrinks. The three lattices shown realize this ratio as $r_1^{(1)}/N_x^{(1)}=2/4$ (left), $r_1^{(2)}/N_x^{(2)}=(2\times5/4)/5$ (center), and $r_1^{(3)}/N_x^{(3)}=(2\times6/4)/6$ (right), illustrating how the continuum limit is approached at shrinking physical transverse extent.}
    \label{fig:constantscale}
\end{figure}

\begin{figure}[!htbp]
\centering
    \includegraphics[width=\linewidth]{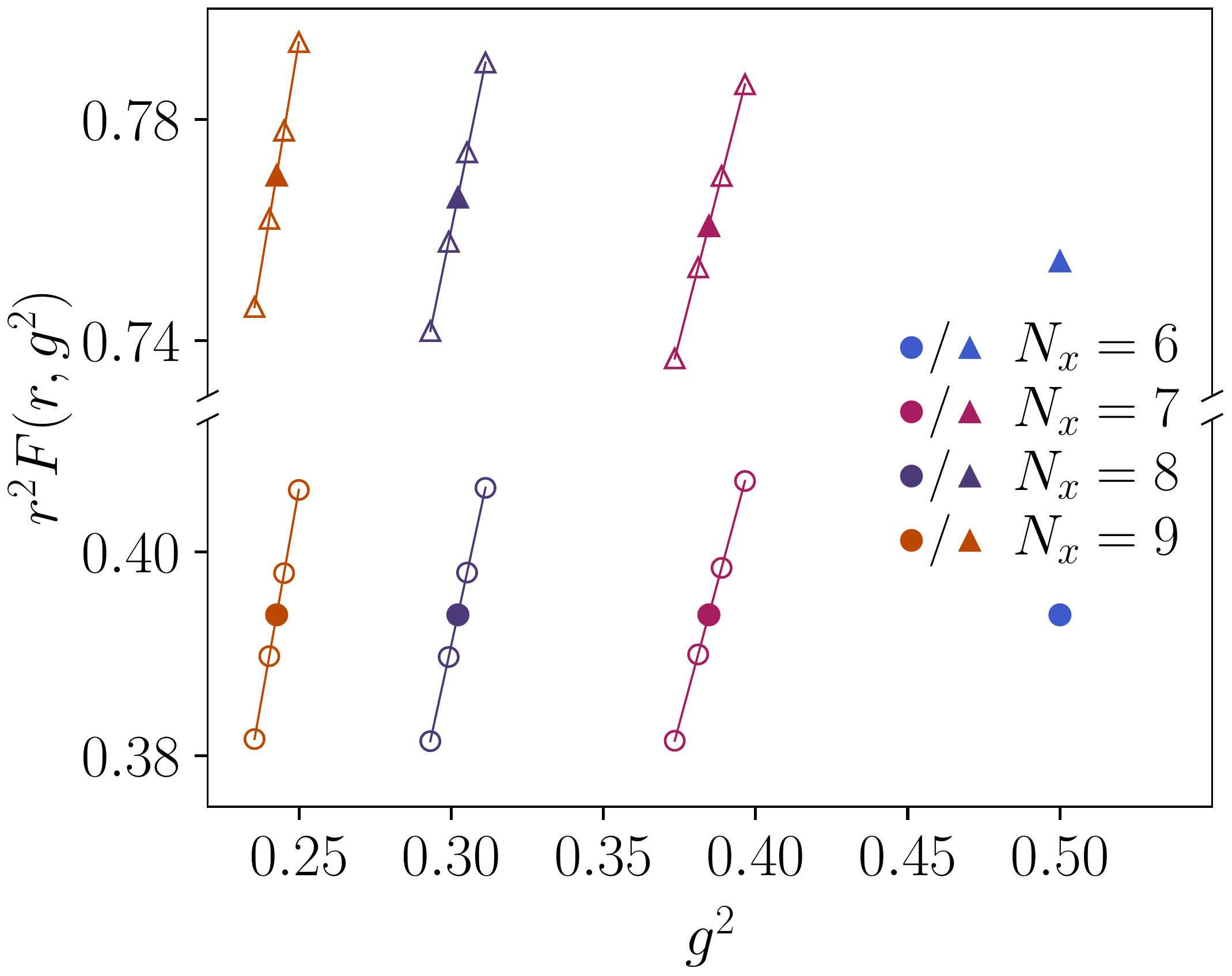}
    \caption{Tuning of the bare coupling for the illustrative sequence $N_x^{(i)}\in\{6,7,8,9\}$ at $N_y=3$, $g^{2}_{(1)}=0.5$, $s=\sqrt{2}$ and with truncation $\Lambda=1$. The sequence runs from right to left, from $N_x=6$ at $g^2=0.5$ to $N_x=9$ at the smallest $g^2$. Empty markers are the values obtained from simulations, from which a quadratic curve is fitted; from the curve the reference values (filled markers) are interpolated. The lower cluster shows the matching condition at the reference scale $r_1$; the upper cluster shows the corresponding values of the step-scaling observable.}
    \label{fig:tuning}
\end{figure}

We now illustrate the finite-volume continuum sequence and tuning described in Sec.~\ref{sec:theory_FVS} with a concrete example. The sequence is built by increasing $N_x$ at fixed $N_y=3$, which keeps the ratio $r_1^{(i)}/N_x^{(i)}$ fixed while the transverse physical size $a N_y$ shrinks as $a\to 0$ (Fig.~\ref{fig:constantscale}).

Starting from a $6\times 3$ lattice at $g^{2}_{(1)}=0.5$ with $r_1=2.0$ in lattice units, $s=\sqrt{2}$, and $\Lambda=1$, we follow the sequence $N_x^{(i)}\in\{6,7,8,9\}$, a subset of the sequence used for the continuum extrapolation in Sec.~\ref{sec:results_clss}. On each lattice, the bare coupling is tuned so that $\overline{g}^2\left(r_1^{(i)}\right)$ takes the common reference value $u$ of Eq.~\eqref{eq:matching}, by interpolating the measured $\overline{g}^2\left(r_1^{(i)};g^2\right)$ with a second-order polynomial in $g^2$. The resulting matched sequence is shown in Fig.~\ref{fig:tuning}.

\begin{figure}[!ht]
\centering
    \includegraphics[width=\linewidth]{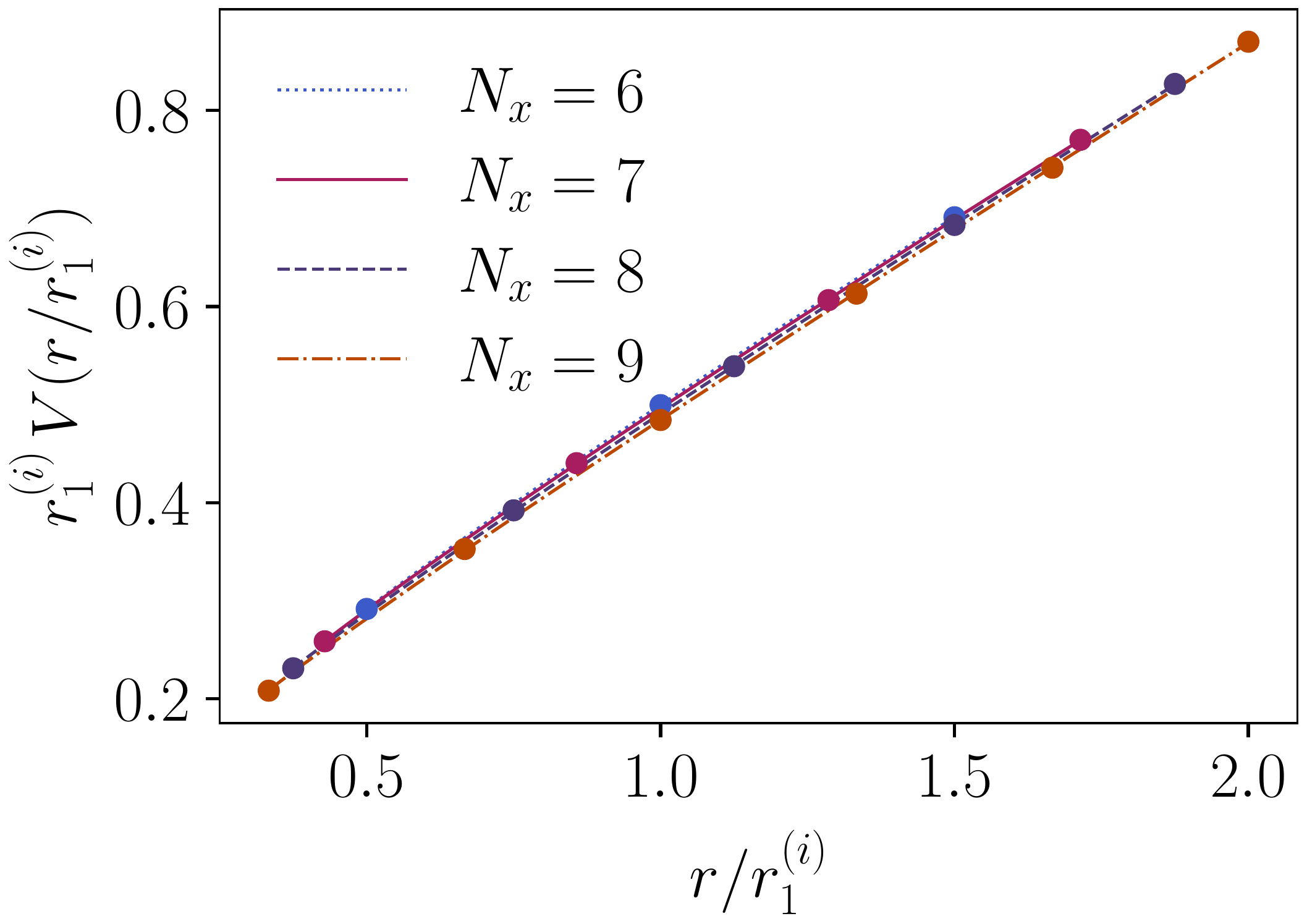}
    \caption{Universality check: dimensionless potential $r_1\,V\left(r\right)$ as a function of $r/r_1$ along the tuned sequence at truncation $\Lambda=1$. Data from different $N_x^{(i)}$ collapse onto a single curve up to small lattice artifacts.}
    \label{fig:universality}
\end{figure}

As a consistency check, we plot the dimensionless potential $r_1\,V\left(r\right)$ as a function of $r/r_1$ for the tuned sequence (Fig.~\ref{fig:universality}). The data collapse onto a single curve up to small lattice artifacts, confirming that the lattices have been matched to the same physical scale.

\subsection{Continuum limit}\label{sec:results_clss}

\begin{figure}[!htbp]
\centering
    \includegraphics[width=\linewidth]{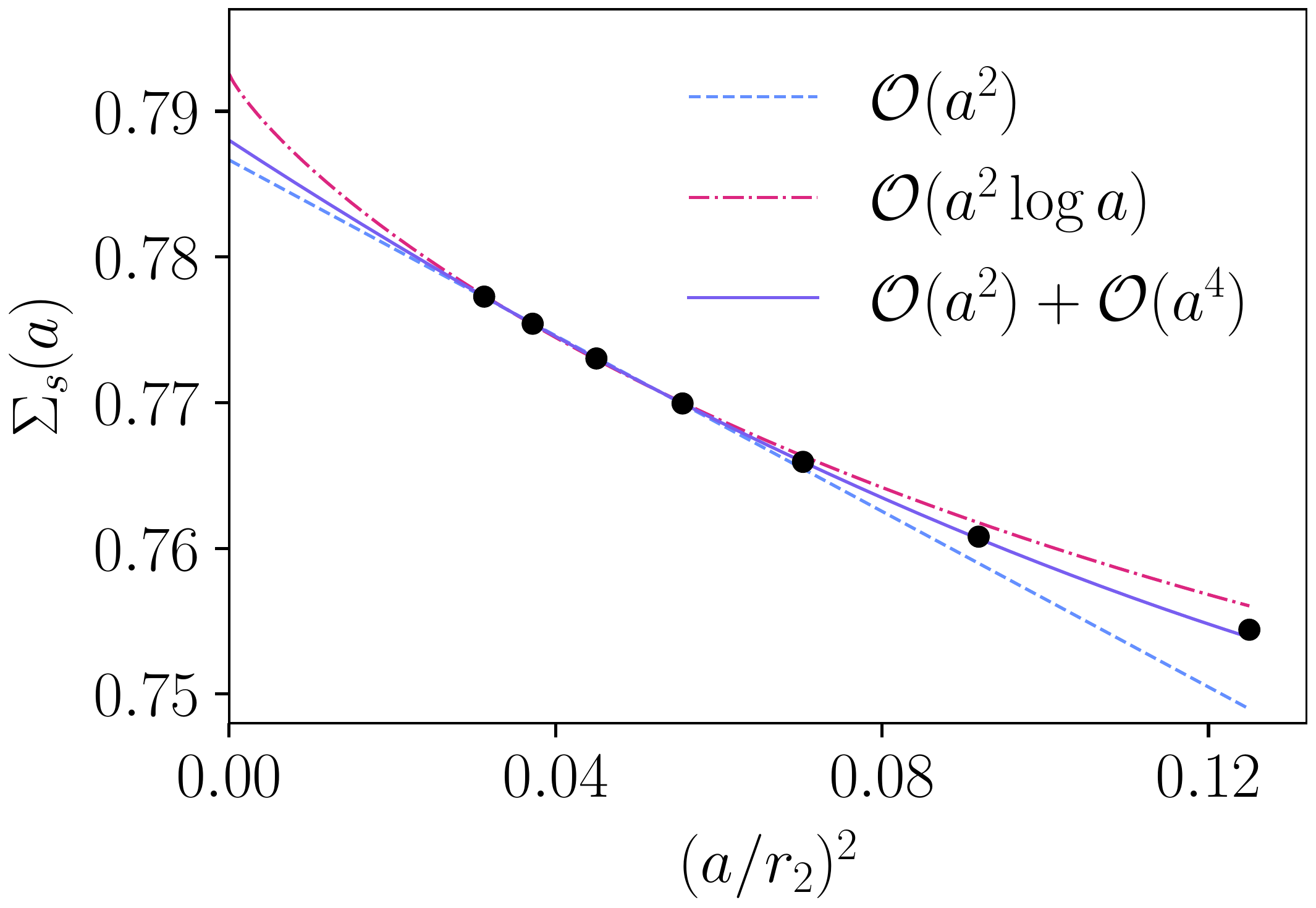}
    \caption{Continuum limit extrapolation of the lattice step-scaling function $\Sigma_{s}$, performed at a truncation level $\Lambda=1$.}
    \label{fig:CLSS}
\end{figure}

We now describe the procedure used to perform the continuum extrapolation of the step-scaling function in the finite-volume scheme. For this, we first measure $\overline{g}^2\left(r_1^{(i)}\right)$ and $\overline{g}^2\left(r_2^{(i)}\right)$ along the sequence at scales $r_1^{(1)} = 2.0$ and $r_2^{(1)} = s\,r_1^{(1)} = 2\sqrt{2}$ (in lattice units), keeping $N_y=3$ and $\Lambda=1$ fixed and starting from $g^{2}_{(1)}=0.5$ on $N_x=6$. The sequence is extended to $N_x^{(i)} \in \{7,8,9,10,11,12\}$ following the finite-volume scheme. Since $r_{2,\text{ph}}$ is fixed by construction, the lattice quantity $r_2^{(i)}=r_{2,\text{ph}} / a^{(i)}$ serves as a convenient proxy for $1/a$. The resulting $\Sigma_s\left(a\right)$ values are shown as black circles in Fig.~\ref{fig:CLSS}; the leftmost point, at $N_x^{(i)}=12$, corresponds to the bare Hamiltonian parameter $g^2\simeq 0.140$.

The $N_x=16,20$ points are presented in Appendix~\ref{app:continuum_extension} and are excluded from the main-text fits.

To extrapolate $\Sigma_s$ to the continuum limit $a\to 0$, we use the perturbative analysis of Ref.~\cite{Luscher:1991wu}, which predicts at $l$-th order
\begin{equation}
    \Sigma_{s}^{(l)}-\sigma_s^{(l)}=\mathcal{O}\!\left(a^2\left(\log a\right)^{l+1}\right)\,,
\end{equation}
up to subleading corrections in even powers of $a$. Motivated by this structure, we compare three fit ansätze
\begin{align}\label{eq:fit3}
    f_{1}\left(a\right) &= c_1 + c_2\,a^2\,, \\
    f_{2}\left(a\right) &= c_1 + c_2\,a^2\log a\,, \\
    f_{3}\left(a\right) &= c_1 + c_2\,a^2 + c_3\,a^{4}\,.\label{eq:fit_a2_a4}
\end{align}
The resulting fits are shown in Fig.~\ref{fig:CLSS}. The four finest data points exhibit clear $\mathcal{O}(a^2)$ scaling and are well described by $f_1$. Including higher-order $\mathcal{O}(a^4)$ corrections via $f_3$ extends the fit range to coarser lattice spacings. In contrast, allowing for a logarithmic correction in $f_2$ leads to a continuum estimate that is more sensitive to the choice of included lattice-spacing range. The logarithmic term does not appear to be dominant in the range of lattice spacings considered.

Quantitatively, the matched reference coupling is $u=0.39382$ with
$s=\sqrt{2}$, giving the infinite-volume tree-level value
$s u=0.55695$. The continuum intercepts are
$f_1(0)=0.78665$, $f_2(0)=0.79256$, and
$f_3(0)=0.78801$, a $41.2\%$--$42.3\%$ increase. This excess is due to a combination of systematic errors and the $s^2$ scaling induced by the linear term in the potential.

In Eq.~\eqref{eq:potential_fit}, $\sigma=\sigma_{\rm phys}a^2$ is
dimensionless. Along the $\Lambda=1$, $N_y=3$ sequence, $\sigma$ and $\delta$
decrease as $g_{\rm lat}^2$ is lowered, while
$2N_y\sigma/g_{\rm lat}^2\simeq1$. The data do not allow us to isolate a monopole contribution to the string tension because finite-volume effects are important in this regime (Appendix~\ref{app:weak_lambda}).

\subsection{Truncation error}\label{sec:results_trunc}

\begin{figure}[!htbp]
\centering
    \includegraphics[width=\linewidth]{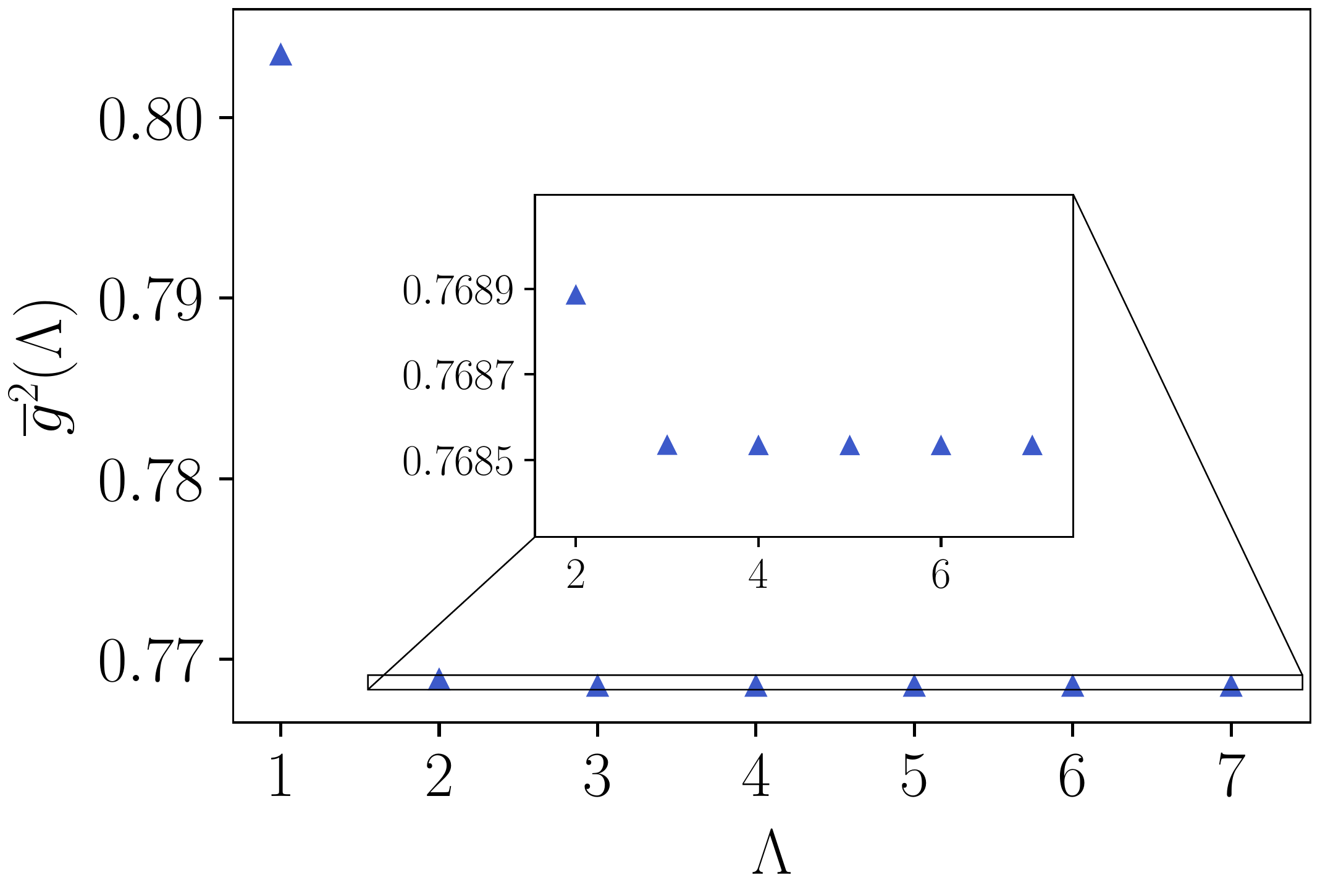}
    \caption{Plot of the renormalized coupling $\overline{g}^2\left(r_1=2.0\right)$ on a $4\times 3$ lattice at $g^2=0.5$ as a function of the truncation $\Lambda\in\{1,\ldots,7\}$. $\overline{g}^2$ approaches a stable value as $\Lambda$ increases.}
    \label{fig:sys_l}
\end{figure}

To assess the bias introduced by the basis truncation, we study how the renormalized coupling $\overline{g}^2$ varies with the local-basis dimension $d_{\mathcal{O}}=2\Lambda+1$ on small lattices. Figure~\ref{fig:sys_l} shows $\overline{g}^2\left(r_1=2.0\right)$ on a $4\times 3$ lattice at $g^2=0.5$ for $\Lambda\in\{1,\ldots,7\}$. The results show a clear convergence pattern, with $\overline{g}^2$ approaching a stable value as $\Lambda$ increases.

\begin{figure}[!htbp]
\centering
    \includegraphics[width=\linewidth]{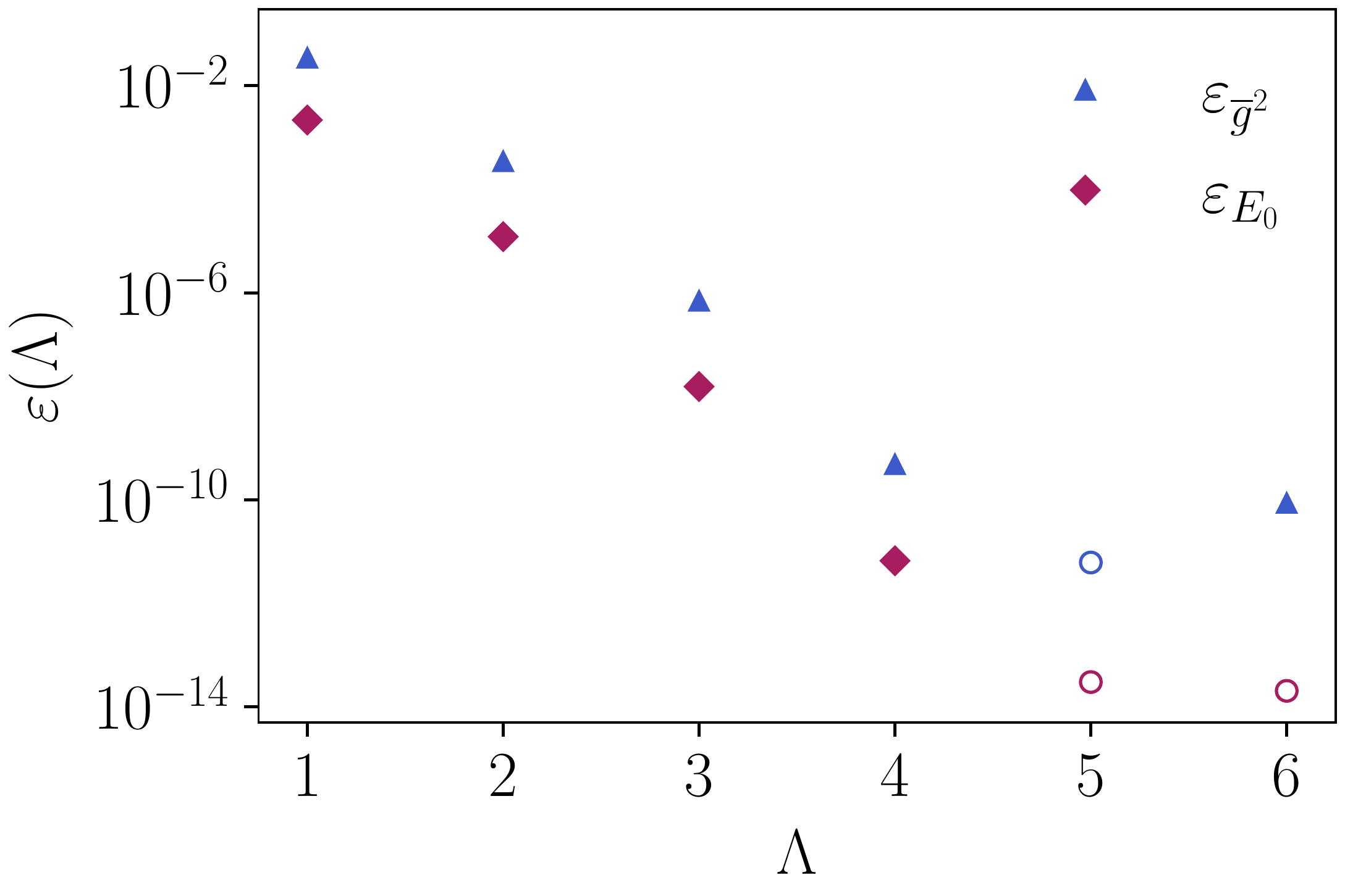}
\caption{Absolute value of the subtracted renormalized coupling $\varepsilon_{\overline{g}^2}\left(\Lambda\right)$ on a $4\times 3$ lattice at $g^2=0.5$ and $r_1=2.0$ as a function of the truncation $\Lambda$ (blue triangles). We also show the convergence of the ground-state energy through $|\varepsilon_{E_0}\left(\Lambda\right)|$ (red diamonds), which reaches machine precision ($\sim 10^{-14}$) for $\Lambda\geq 5$. Negative values are indicated by empty markers.}
    \label{fig:eps_l_log}
\end{figure}

In order to quantify the convergence rate more precisely, we define the subtracted renormalized coupling
\begin{equation}
    \varepsilon_{\overline{g}^2}\left(\Lambda\right)=\overline{g}^2\left(\Lambda\right)-\overline{g}^2\left(\Lambda=7\right)\,,\qquad \Lambda \in \{1,\ldots,6\}\,,
\end{equation}
taking $\overline{g}^2\left(\Lambda=7\right)$ as the converged reference. Figure~\ref{fig:eps_l_log} plots $|\varepsilon_{\overline{g}^2}\left(\Lambda\right)|$ on a logarithmic axis, illustrating an approximately exponential approach to the converged value. Additionally, we track the change in ground-state energy $E_0$ with increasing truncation
\begin{equation}
    \varepsilon_{E_0}\left(\Lambda\right)=E_0\left(\Lambda\right)-E_0\left(\Lambda=7\right)\,,\qquad \Lambda \in \{1,\ldots,6\}\,,
\end{equation}
and plot its absolute value in Fig.~\ref{fig:eps_l_log}. The energy difference falls off exponentially and reaches the level of double-precision round-off ($\sim 10^{-14}$) by $\Lambda=5$, demonstrating convergence to machine precision; for $\Lambda\geq 5$ negative values are consistent with numerical noise (empty markers).

\begin{figure}[!htbp]
\centering
    \includegraphics[width=\linewidth]{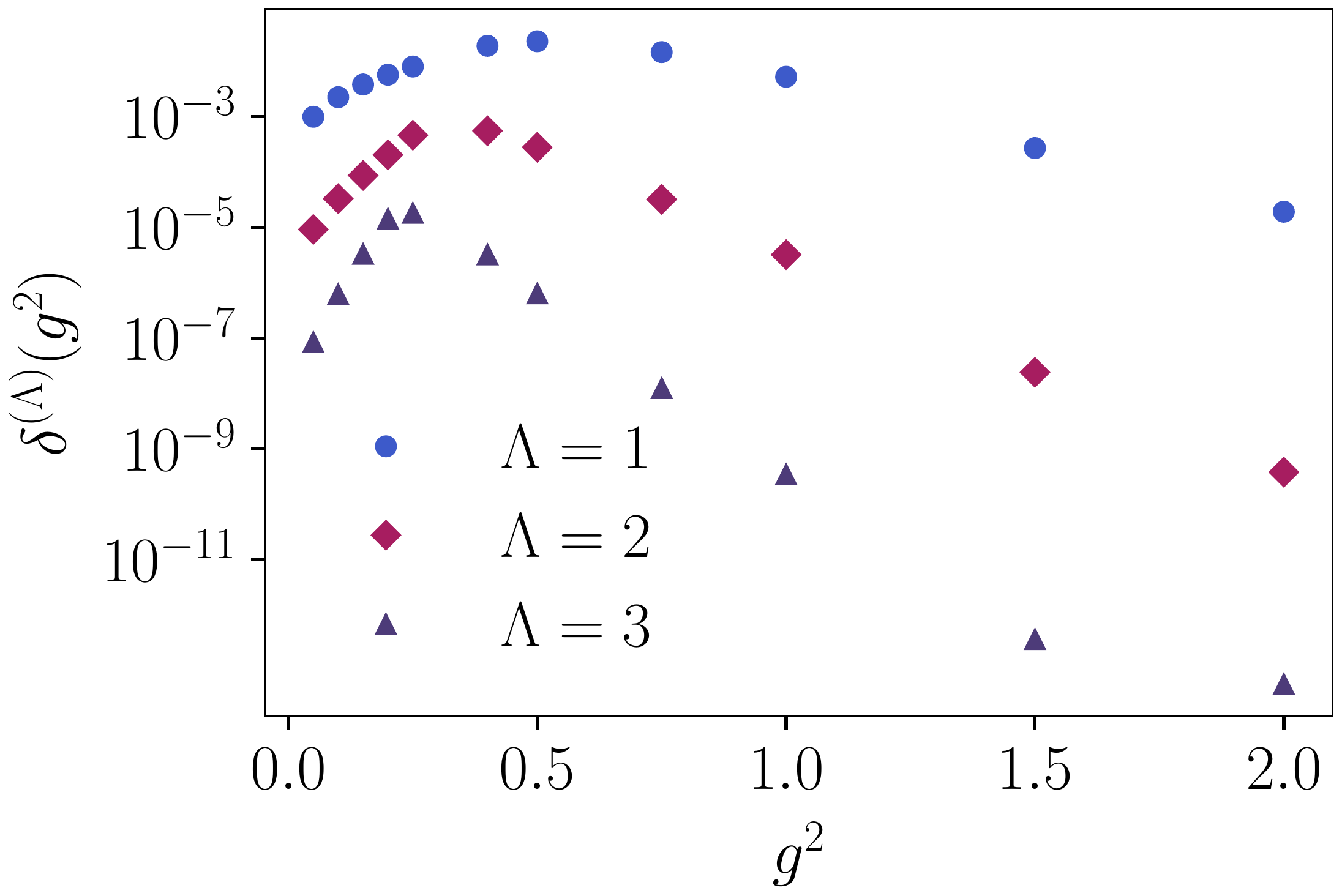}
    \caption{Discrete difference $\delta^{(\Lambda)}\left(g^2\right)=\overline{g}^{2(\Lambda)}\left(g^2\right)-\overline{g}^{2(\Lambda+1)}\left(g^2\right)$ of the renormalized coupling between two successive truncation levels, shown on a $6\times 3$ lattice at $r_1=2.0$ across the weak- to strong-coupling range for $\Lambda\in\{1,\ldots,4\}$. As $\Lambda$ increases, the maximum shifts toward smaller values of $g^2$.}
    \label{fig:g_vs_l_rel}
\end{figure}

Finally, to probe the coupling dependence of the truncation error, we define the discrete difference
\begin{equation}
    \delta^{(\Lambda)}\left(g^2\right)= \overline{g}^{2(\Lambda)}\left(g^2\right)-\overline{g}^{2(\Lambda+1)}\left(g^2\right)\,,
\end{equation}
i.e.\ the discrete difference of the renormalized coupling between two successive truncation levels, or equivalently the change in $\overline{g}^2$ when the local basis is enlarged by two states. We scan $\delta^{(\Lambda)}\left(g^2\right)$ across the weak- to strong-coupling range for each $\Lambda\in\{1,\ldots,4\}$ on a $6\times 3$ lattice at $r_1=2.0$ (Fig.~\ref{fig:g_vs_l_rel}). The truncation effect peaks at intermediate coupling and becomes smaller in both the strong- and weak-coupling regimes. As $\Lambda$ increases, the peak shifts toward smaller values of $g^2$.

At $g^2=0.01$ on a $4\times3$ lattice, $\overline{g}^2(r/a=2)$ at $\Lambda=2$ differs from the $\Lambda=7$ reference by $4.42\times10^{-5}$ in relative terms. For the tested $\Lambda\geq4$, the remaining differences are below numerical resolution (Appendix~\ref{app:weak_lambda}).

\subsection{Finite transverse-size effects and the large-\texorpdfstring{$N_y$}{Ny} limit}\label{sec:results_finitesize}

\begin{figure}[!htbp]
\centering
\includegraphics[width=\linewidth]{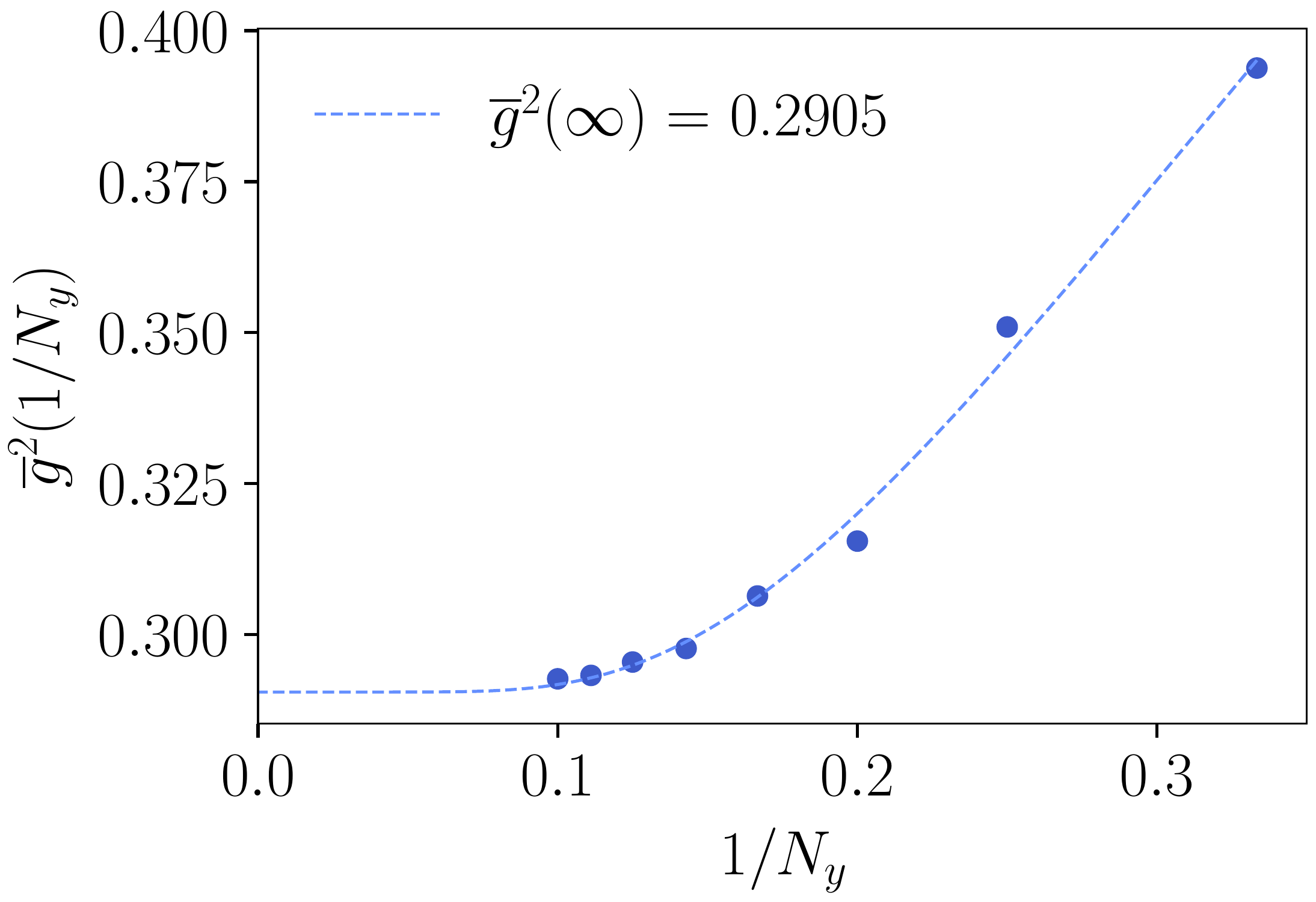}
\caption{Dependence of the renormalized coupling $\overline{g}^2\left(r_1=2.0\right)$ on the transverse size $N_y$. The curve is an exponential extrapolation to the large-$N_y$ limit, with asymptote $\overline{g}^2\left(\infty\right)$.}
\label{fig:nyeffects}
\end{figure}

To assess the impact of the finite transverse lattice extent, we measure the renormalized coupling $\overline{g}^2\left(r_1=2.0\right)$ on lattices of size $6\times N_y$, with $N_y\in\{3,\ldots,10\}$. The results are shown in Fig.~\ref{fig:nyeffects}. We model the approach to the large-$N_y$ limit with the exponential ansatz
\begin{equation}
    f\left(N_y\right)=\overline{g}^2\left(\infty\right)+A\,\exp\!\bigl[-\kappa\,\left(N_y-N_0\right)\bigr].
\end{equation}
Here, $\overline{g}^2\left(\infty\right)$ denotes the large-$N_y$ limit, $A$ the size of the finite-$N_y$ correction, $\kappa$ its decay rate, and $N_0$ an effective offset. The data show a clear convergence toward an asymptotic value as $N_y$ increases, with an approximately monotonic behavior toward $\overline{g}^2\left(\infty\right)$. Small deviations from the fit are visible for $N_y\geq 6$, i.e.\ once the lattice becomes at least as large in the transverse direction as in the longitudinal one.

\begin{figure}[t]
\centering
\includegraphics[width=\linewidth]{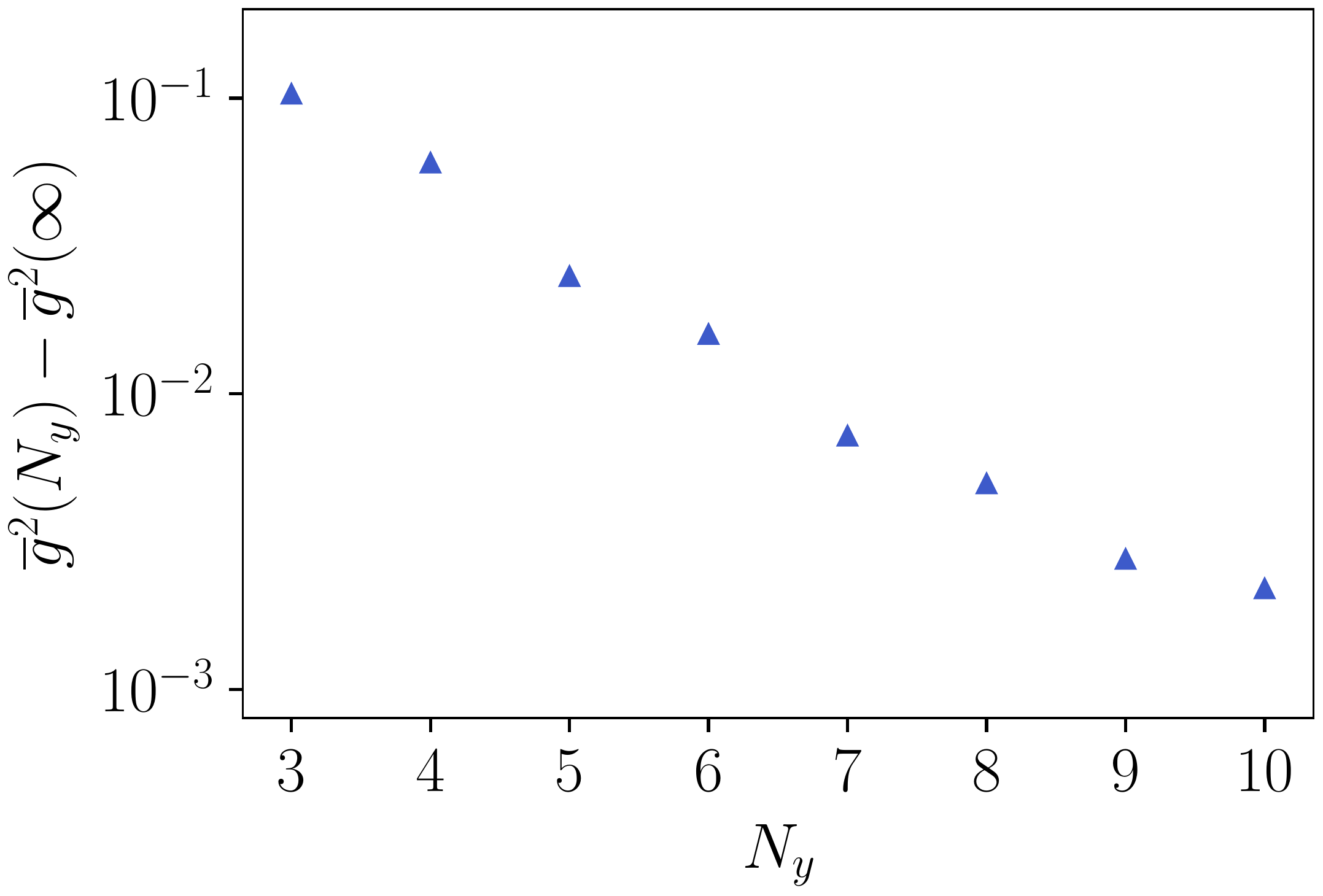}
    \caption{Finite-$N_y$ correction to the renormalized coupling, $\overline{g}^2\left(N_y\right)-\overline{g}^2\left(\infty\right)$, for $\overline{g}^2\left(r_1=2.0\right)$ on $6\times N_y$ lattices. The logarithmic scale highlights the exponential convergence toward the large-$N_y$ limit.}
\label{fig:nysub}
\end{figure}

To quantify the finite-$N_y$ effect, in Fig.~\ref{fig:nysub} we plot the difference
\begin{equation}
    \overline{g}^2\left(N_{y}\right)-\overline{g}^2\left(\infty\right)
\end{equation}
on a logarithmic scale, which highlights the exponential nature of the convergence. This representation makes the role of $\overline{g}^2\left(\infty\right)$ explicit, as it is the reference value used to define the finite-$N_y$ correction and estimate the associated systematic.

\section{Discussion: continuum limit and systematic effects}\label{sec:discussion}

The central methodological contribution of this work is the finite-volume scheme introduced in Sec.~\ref{sec:theory_FVS}, which fixes the ratio $r_1^{(i)}/N_x^{(i)}$ along the continuum sequence so that the physical reference scale is identical on each lattice. The dimensionless potential $r_1\,V\left(r\right)$ collapses onto a single curve across the tuned sequence (Fig.~\ref{fig:universality}), confirming that the matched lattices realize the same physical scale; Appendix~\ref{app:uncontrolled} provides the corresponding negative control, where relaxing the fixed-ratio prescription produces a non-monotonic $\Sigma_s\left(a\right)$ in which finite-volume effects dominate.

Our procedure yields a stable extrapolation of the fixed-$N_y$ sequence: holding $r_1/N_x$ fixed controls the longitudinal geometry, but $N_y=3$ leaves transverse finite-volume and open-boundary effects because $aN_y$ shrinks along the sequence.

Since $g_{\mathrm{cont}}^2$ is dimensionful, part of the scale dependence of $\overline{g}^2(r)$ is already present at tree level and should not be interpreted as quantum running. Within our fit ansatz, a nonzero fitted linear coefficient produces scaling beyond $s u$. The observed $41.2\%$--$42.3\%$ excess is therefore a departure from the infinite-volume tree-level dimensional baseline, but it cannot yet be identified uniquely with a continuum nonperturbative contribution. It can also contain finite-volume and open-boundary effects, $\Lambda=1$ projection effects and extrapolation systematics.

Appendix~\ref{app:weak_lambda} compares the $\Lambda=1,2$ sequences at $N_y=3$ and adds a $\Lambda=2$ sequence at $N_y=5$, including $N_x=16,20$. At $N_y=3$, increasing $\Lambda$ from 1 to 2 changes the coarser points more than the $N_x\leq12$ extrapolated intercepts, which differ by less than $0.2\%$.

The present study should be regarded as a proof-of-concept calculation rather than a precision determination: in quantitative applications, both local-basis truncation effects and finite-size effects must be explicitly controlled and propagated as systematic uncertainties, since they give sizeable contributions on the lattices accessible here.

The force scheme also ties several systematic effects to the use of an extended charged object. At finite local-basis truncation, the background field $n^{(\gamma)}$ introduces a dependence on the path connecting the charges, extracting the force requires a functional fit of the potential and the transverse spread of the flux tube makes the result sensitive to the transverse lattice extent.

At $\Lambda=1$, a one-plaquette dent raises the energy by $4.6\%$--$7.1\%$ of the straight-path potential. Since increasing $\Lambda$ lowers each variational energy toward the common untruncated value, the straight path is the preferred finite-$\Lambda$ representation among those tested. At $\Lambda=3$, the dent residual is reduced by about $99\%$. Appendix~\ref{app:path_deformation} gives the comparison and its limitations.

A possible implementation of a local scheme is to use the plaquette expectation value, as discussed in Ref.~\cite{Clemente:2022cka}. If validated as a renormalized coupling with a continuum-matching and step-scaling prescription for this local Hamiltonian observable, it would remove the external charges, background path, and potential fit, but not automatically finite-volume or truncation effects. We do not adopt it because that validation is not yet available.

On the tested small lattices and observables, the cutoff scan shows an approximately exponential approach to a stable value. At $g^2=0.5$ a truncation level of $\Lambda=4$ is needed to reach a precision comparable to the SVD cutoff. Beyond this, $\varepsilon_{\overline{g}^2}\left(\Lambda\right)$ fluctuates around a stable value; the sign change at $\Lambda=5$ in Fig.~\ref{fig:eps_l_log} indicates that the remaining variations are at the level of numerical precision.

We observe that truncation effects are largest in an intermediate-coupling region, while they are reduced in both the strong-coupling and weak-coupling regimes. For $g^2\gtrsim 1$--$2$, enlarging the local plaquette basis has only a limited impact on the measured renormalized coupling, as the plaquette eigenstates converge to the electric-flux eigenstates in the large coupling limit. As $g^2$ is lowered, $\delta^{(\Lambda)}$ grows and reaches a maximum in the crossover region, where both the electric and the magnetic parts of the dual Hamiltonian contribute comparably. This regime is therefore most sensitive to the size of the truncated local basis.

Toward weak coupling, as $\Lambda$ increases, the maximum of
$\delta^{(\Lambda)}$ shifts toward smaller $g^2$. This may reflect improved
resolution of the local basis, but it could also signal a $\Lambda$-dependent
crossover in the projected theory. An analogous shift occurs when compact
$\Uone$ is approximated by a $\mathbb{Z}_N$ gauge theory, where increasing $N$ moves
the transition toward weaker coupling~\cite{Borisenko:2013qza}. This analogy
is only qualitative, as our finite Mathieu basis is not a $\mathbb{Z}_N$ gauge
theory, so the observed peak shift does not identify the projected theory with
one. The basis cutoff must be removed at each lattice spacing before
taking $a\to0$, or increased along the sequence with verified
convergence.
Appendix~\ref{app:weak_lambda} discusses this comparison.
It compares the fitted string
coefficient with the expected semiclassical monopole scaling, explains how
the finite transverse width affects this comparison, and presents a basis convergence test at weak coupling.

Finally, we emphasize that the finite transverse-size effects approach their $N_y\to\infty$ value exponentially in $N_y$ (Fig.~\ref{fig:nysub}). This is to be expected, as the string becomes more insensitive to the boundary regions as $N_y$ increases. Nonetheless, working at a finite transverse size has a significant impact on the running coupling, and the extrapolated value is about $26\%$ smaller than the value at $N_y=3$, used for the lattice-spacing extrapolation. The scheme fixes the \emph{longitudinal} ratio $r_1^{(i)}/N_x^{(i)}$, rather than the transverse physical extent, so the transverse physical size in units of $r_1$ shrinks along the sequence. The limit we reach is thus that of a fixed transverse geometry (effectively a thin strip) rather than the two-dimensional theory, and the $N_y\to\infty$ correction remains a separate systematic. It does not, however, spoil the extrapolation: the matched sequence follows a clean $a^2$ approach (Fig.~\ref{fig:CLSS}) with no resolvable lower-order term from the shrinking width, which only sets the (strip) value to which the fit converges. A fully two-dimensional result would require $N_y\to\infty$ at each lattice spacing, beyond the reach of the MPS ansatz used here.

\section{Conclusion \& Outlook}\label{sec:conclusion}

In this work we have developed and tested a finite-volume scheme to perform the continuum extrapolation of the lattice step-scaling function in the Hamiltonian formulation of compact $\Uone$ lattice gauge theory in (2+1) dimensions.
Building on the running-coupling strategy proposed in Refs.~\cite{Clemente:2022cka, Crippa:2024cqr}, we obtain a stable extrapolation at fixed $\Lambda=1$ and $N_y=3$ by holding the longitudinal ratio $r_1/N_x$ fixed. The resulting sequence is smooth on the small volumes accessible to Hamiltonian and prospective quantum simulations and avoids the non-monotonic behavior seen without the fixed-ratio prescription.

To this end, we have combined the dual Hamiltonian formulation with the plaquette-state operator basis introduced in Ref.~\cite{Jakobs:2025zcv}, which allows us to reach the weak-coupling regime relevant for continuum extrapolations. We have extended the dual formulation by including static charges and derived the corresponding Hamiltonian explicitly in App.~\ref{sec:app}. This has made it possible to study directly the formation of a confining flux tube, to extract the static potential and the corresponding force, and from these to define a force-based renormalized coupling.

We have also quantified the sensitivity to the local plaquette-basis
cutoff and the transverse lattice size. The extrapolation is performed
at fixed $\Lambda$ and $N_y$, so the transverse physical width $aN_y$
shrinks along the sequence. Identifying the extrapolated value with
the continuum step-scaling function of the untruncated compact
$\Uone$ theory therefore still requires $\Lambda$ convergence and
control of the transverse physical extent. Reaching larger $N_y$
at comparable precision remains the main practical limitation of
the current column-major MPS formulation.

Even though Hamiltonian tensor-network simulations are not yet competitive with Monte Carlo methods in terms of accessible volumes, they offer several structural advantages~\cite{Banuls:2019rao,Banuls:2019bmf}, most notably the absence of both statistical sampling noise and autocorrelation, which is particularly relevant for the present work.

A natural next step will be the use of genuinely two-dimensional tensor-network methods, such as PEPS~\cite{Verstraete:2004cf,Orus:2013kga}, isoTNs~\cite{Zaletel:2020roc}, or augmented TTNs~\cite{Felser:2020ntd}, in order to reduce the dominant finite-size systematic and access larger transverse volumes. The present construction provides, on a longer timescale, a concrete step toward running-coupling studies in Hamiltonian lattice gauge theory and toward future implementations on quantum devices in the NISQ era~\cite{Preskill:2018jim,DiMeglio:2023nsa}. Conceptually, it will also be interesting to extend the present strategy to more general gauge theories, including $\Uone$ with dynamical matter~\cite{Banuls:2013jaa,Buyens:2013yza}, non-Abelian gauge groups~\cite{Silvi:2014pta,Rico:2013qya}, and higher dimensions~\cite{Felser:2019xyv,Magnifico:2020bqt}. More broadly, the framework developed here opens the possibility of studying step-scaling and continuum extrapolations~\cite{Luscher:1991wu}  within a unified Hamiltonian setting.

\begin{acknowledgments}
  This project was funded by the Deutsche Forschungsgemeinschaft (DFG,
  German Research Foundation) as part of the CRC 1639 NuMeriQS --
  project no.\ 511713970 and under Germany's Excellence Strategy --
  Cluster of Excellence ``Color meets Flavor'' (CmF) EXC 3107 --
  533766364 and Cluster of Excellence ``Matter and Light for Quantum
  Computing'' (ML4Q) EXC 2004/2 -- 390534769.
  The authors gratefully acknowledge access to the Marvin cluster of the University of Bonn.
  We thank Arianna Crippa for her contribution in an early stage and for fruitful discussions. We thank Andrea Bulgarelli, Christiane Franziska Groß, Simone Romiti, Paolo Stornati, Luca Tagliacozzo and Lorenzo Verzichelli for helpful comments.

\end{acknowledgments}

\bibliography{bibliography}

\begin{appendix}
  \raggedbottom
  \section{Deriving the Hamiltonian with static charges}\label{sec:app}

The Hamiltonian in Eq.~\eqref{eq:dual_hamiltonian_sc} can be derived using the transfer matrix formalism as in Ref.~\cite{Unmuth-Yockey:2018xak}.
Following their approach, one first constructs the partition function and then identifies the corresponding transfer matrix. The main difference here is the inclusion of static charges in the system.

\subsection{The partition function with static charges}

\begin{figure}[!t]
    \centering
    \resizebox{\linewidth}{!}{%
    \begin{tikzpicture}[
  scale=1.0,
  x={(1cm,0cm)},
  y={({cos(60)*0.7cm},{sin(60)*-0.4cm})},
  z={(0cm,1.0cm)},
  site/.style={circle, fill=black!50, draw=black!70, inner sep=0pt, minimum size=5pt},
  link/.style={thick, gray!60},
  polyakov/.style={ultra thick, line cap=round,
    postaction={decorate, decoration={
      markings,
      mark=at position 0.45 with {\arrow[scale=1.2]{Stealth}}
    }}
  },
]

\def\tmin{0}
\def\tmax{2}

\draw[-Stealth, thick, black!80] (-0.35,-0.35,0) -- (3.2,-0.35,0)
  node[below, font=\small] {$x$};
\draw[-Stealth, thick, black!80] (-0.35,-0.35,0) -- (-0.35,3.2,0)
  node[below right, font=\small] {$y$};
\draw[-Stealth, thick, black!80] (-0.35,-0.35,0) -- (-0.35,-0.35,3.0)
  node[above, font=\small] {$\tau$};

\foreach \t in {0,1,2}{

  \foreach \yy in {0,1,2}{
    \foreach \xx in {0,1}{
      \draw[link] (\xx,\yy,\t) -- (\xx+1,\yy,\t);
    }
  }

  \foreach \xx in {0,1,2}{
    \foreach \yy in {0,1}{
      \draw[link] (\xx,\yy,\t) -- (\xx,\yy+1,\t);
    }
  }
}

\foreach \xx in {0,1,2}{
  \foreach \yy in {0,1,2}{
    \ifnum\xx=0
      \ifnum\yy=0
      \else
        \foreach \t in {0,1}{
          \draw[link] (\xx,\yy,\t) -- (\xx,\yy,\t+1);
        }
      \fi
    \else
      \ifnum\xx=2
        \ifnum\yy=2
        \else
          \foreach \t in {0,1}{
            \draw[link] (\xx,\yy,\t) -- (\xx,\yy,\t+1);
          }
        \fi
      \else
        \foreach \t in {0,1}{
          \draw[link] (\xx,\yy,\t) -- (\xx,\yy,\t+1);
        }
      \fi
    \fi
  }
}

\foreach \t in {0,1,2}{
  \foreach \xx in {0,1,2}{
    \foreach \yy in {0,1,2}{
      \node[site] at (\xx,\yy,\t) {};
    }
  }
}

\begin{scope}[
  shift={(current bounding box.north east)},
  x={(1cm,0cm)}, y={(0cm,1cm)}, z={(0cm,0cm)}
]
  \node[anchor=north west, align=left, font=\small] at (0.1,-0.1) {
    \textcolor{red}{$\longrightarrow$ Polyakov line ($+q$)}\\[4pt]
    \textcolor{blue}{$\longrightarrow$ Polyakov line ($-q$)}\\[4pt]
    \textcolor{gray}{$\rule{1cm}{0.6pt}$ gauge link $U_\mu$}
  };
\end{scope}

\draw[polyakov, red] (0,0,0) -- (0,0,1);
\draw[polyakov, red] (0,0,1) -- (0,0,2);

\draw[polyakov, blue] (2,2,2) -- (2,2,1);
\draw[polyakov, blue] (2,2,1) -- (2,2,0);

\end{tikzpicture}%
    }
    \caption{Schematic of the lattice setup for two static charges $+q$ and $-q$ at spatial sites $\vec{x}_1$ and $\vec{x}_2$, respectively. The charges trace out Polyakov lines in Euclidean time, which are represented by the red and blue lines. The gray lines represent the gauge links $U_\mu$ on the lattice.}
    \label{fig:SC3D}
\end{figure}
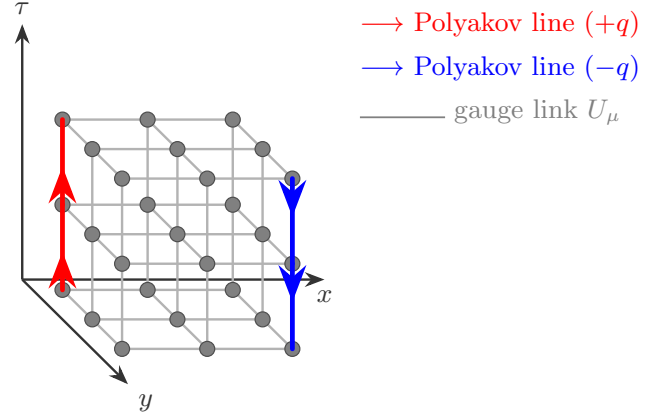

We begin with the Euclidean partition function for compact $\Uone$ gauge theory in the
Wilson formulation
\begin{equation}
Z = \int \mathcal{D}[A_{x,\mu}]\, e^{-S[A]}\, ,\quad
S[A] = -\beta \sum_{x,\mu<\nu} \cos F_{x,\mu\nu}\,,
\end{equation}
where the plaquette field strength is
\begin{equation}
F_{x,\mu\nu}
= \Delta_\mu A_{x,\nu} - \Delta_\nu A_{x,\mu}\,,
\end{equation}
and
\begin{equation}
    \int \mathcal{D}A \equiv \prod_{x,\mu} \int_{-\pi}^{\pi}\frac{dA_{x,\mu}}{2\pi}\,.
\end{equation}
The forward derivative is defined as
\begin{equation}
    \Delta_\mu f(x) = f(x+\hat\mu)-f(x)\,.
\end{equation}
We introduce static charges by coupling the gauge field to a conserved external current
$J_\mu(x)$
\begin{equation}
Z_{q\bar q}
= \int \mathcal{D}[A_{x,\mu}]\;
\exp\!\left[
    -S[A] + i \sum_{x,\mu} J_\mu(x) A_{x,\mu}
\right].
\label{eq:Zqbarq-start}
\end{equation}
For a static charge $+q$ at $\vec{x}_1$ and a static charge $-q$ at $\vec{x}_2$ we choose
\begin{equation}
J_0(\vec{x},t)
    = q\bigl(\delta_{\vec{x},\vec{x}_1}-\delta_{\vec{x},\vec{x}_2}\bigr),
\qquad
J_i(\vec{x},t)=0,
\label{eq:J-static}
\end{equation}
so that each charge traces out a Polyakov line in Euclidean time, which is schematically illustrated in Fig.~\ref{fig:SC3D}.

Using the following identity of the Bessel functions
\begin{equation}
    \exp{\left(a\cos{b}\right)}=\sum_{n\in\mathbb{Z}}I_n(a)\exp{\left( inb \right)}\,,
\end{equation}
the partition function becomes
\begin{equation}
\begin{aligned}
&Z_{q\bar q}
\\[4pt]
&= \int \mathcal D[A_{x,\mu}]\;
\exp\!\left[
\beta\sum_{x,\mu<\nu}\cos F_{x,\mu\nu}
\;+\;
i\sum_{x,\mu} J_\mu(x)\,A_{x,\mu}
\right]
\\[4pt]
&= \int \mathcal D[A_{x,\mu}]\;
\prod_{x,\mu<\nu}\;
\sum_{n_{x,\mu\nu}\in\mathbb Z}
I_{n_{x,\mu\nu}}(\beta)\,
\exp\!\left(i\,n_{x,\mu\nu}\,F_{x,\mu\nu}\right)\;\\
&\quad\times\exp\!\left(i\sum_{x,\mu}J_\mu(x)\,A_{x,\mu}\right)
\\[4pt]
&= \int \mathcal D[A_{x,\mu}]\;
\sum_{\{n_{x,\mu\nu}\in\mathbb Z\}}
\prod_{x,\mu<\nu} I_{n_{x,\mu\nu}}(\beta)\;\\
&\quad\times\exp\!\left[
i\sum_{x,\mu<\nu} n_{x,\mu\nu}
\bigl(\Delta_\mu A_{x,\nu}-\Delta_\nu A_{x,\mu}\bigr)\right.
\;\\
&\quad \left.+\;
i\sum_{x,\mu}J_\mu(x)\,A_{x,\mu}
\right]\,.
\end{aligned}
\end{equation}

Using summation by parts
\begin{equation}
    \sum_{k=0}^{n-1} f_k\Delta g_k = \left(f_{n} g_{n} - f_0 g_0\right) - \sum_{k=0}^{n-1} g_{k+1}\Delta f_k\,,
\end{equation}
one obtains a bulk term in which the derivative acts on $n_{x,\mu\nu}$, plus a boundary contribution. For periodic boundary conditions the boundary term vanishes identically. For open boundary conditions it vanishes only after imposing the corresponding boundary condition on the fields, e.g. by imposing that no external electric flux can enter the system. The result is
\begin{equation}
\begin{aligned}
&\sum_{x,\mu<\nu} n_{x,\mu\nu}
\bigl(\Delta_\mu A_{x,\nu}-\Delta_\nu A_{x,\mu}\bigr)\\
&=
\sum_{x,\mu<\nu}
\Bigl(
- A_{x,\nu}\,\Delta_\mu n_{x,\mu\nu}
+ A_{x,\mu}\,\Delta_\nu n_{x,\mu\nu}
\Bigr)
\\
&=
-\sum_{x,\mu,\nu} A_{x,\mu}\,\Delta_\nu n_{x,\nu\mu}\,,
\end{aligned}
\end{equation}
where $\Delta$ from now on denotes the backward derivative.
We obtain the partition function
\begin{align}
Z_{q\bar q}
&= \int \mathcal D[A_{x,\mu}]\;
\sum_{\{n_{x,\mu\nu}\in\mathbb Z\}}
\prod_{x,\mu<\nu} I_{n_{x,\mu\nu}}(\beta)\;\\
\notag&\quad\times\exp\!\left[
i\sum_{x,\mu} A_{x,\mu}\Bigl(-\sum_\nu \Delta_\nu n_{x,\nu\mu}+J_\mu(x)\Bigr)
\right]\,.
\end{align}
Here, we extend $n_{\mu\nu}$ antisymmetrically to all index pairs, using the relation $n_{\mu\nu}\coloneq -n_{\nu\mu}$\,.

\begin{figure}[!t]
    \centering
    \resizebox{\linewidth}{!}{%
    \begin{tikzpicture}[
  scale=1.0,
  x={(1cm,0cm)},
  y={({cos(60)*0.7cm},{sin(60)*-0.4cm})},
  z={(0cm,1.0cm)},
  site/.style={circle, fill=black!50, draw=black!70, inner sep=0pt, minimum size=5pt},
  link/.style={thick, gray!60},
  polyakov/.style={ultra thick, line cap=round,
    postaction={decorate, decoration={
      markings,
      mark=at position 0.45 with {\arrow[scale=1.2]{Stealth}}
    }}
  },
]

\draw[-Stealth, thick, black!80] (-0.35,-0.35,0) -- (3.2,-0.35,0)
  node[below, font=\small] {$x$};
\draw[-Stealth, thick, black!80] (-0.35,-0.35,0) -- (-0.35,3.2,0)
  node[below right, font=\small] {$y$};
\draw[-Stealth, thick, black!80] (-0.35,-0.35,0) -- (-0.35,-0.35,3.0)
  node[above, font=\small] {$\tau$};

\foreach \t in {0,1,2}{
  \draw[link] (1,0,\t) -- (2,0,\t);
}

\foreach \t in {0,1,2}{
  \draw[link] (2,0,\t) -- (2,1,\t);
}

\foreach \t in {0,1}{
  \draw[link] (2,0,\t) -- (2,0,\t+1);
}

\foreach \t in {0,1,2}{
  \node[site] at (2,0,\t) {};
}

\fill[yellow!60, opacity=0.45]
  (0,0,0) -- (1,0,0) -- (1,0,2) -- (0,0,2) -- cycle;
\foreach \frac in {0.15, 0.35, 0.55, 0.75, 0.95}{
  \draw[yellow!30!orange, opacity=0.9, thick]
    ({0+\frac*1},{0},{0}) -- ({0+\frac*1},{0},{2});
}

\fill[yellow!60, opacity=0.45]
  (1,0,0) -- (1,1,0) -- (1,1,2) -- (1,0,2) -- cycle;
\foreach \frac in {0.15, 0.35, 0.55, 0.75, 0.95}{
  \draw[yellow!60!green, opacity=0.9, thick]
    ({1},{0+\frac*1},{0}) -- ({1},{0+\frac*1},{2});
}

\fill[yellow!60, opacity=0.45]
  (1,1,0) -- (2,1,0) -- (2,1,2) -- (1,1,2) -- cycle;
\foreach \frac in {0.15, 0.35, 0.55, 0.75, 0.95}{
  \draw[yellow!30!orange, opacity=0.9, thick]
    ({1+\frac*1},{1},{0}) -- ({1+\frac*1},{1},{2});
}

\fill[yellow!60, opacity=0.45]
  (2,1,0) -- (2,2,0) -- (2,2,2) -- (2,1,2) -- cycle;
\foreach \frac in {0.15, 0.35, 0.55, 0.75, 0.95}{
  \draw[yellow!60!green, opacity=0.9, thick]
    ({2},{1+\frac*1},{0}) -- ({2},{1+\frac*1},{2});
}

\draw[yellow!40!black, thick, dashed]
  (0,0,0) -- (1,0,0) -- (1,1,0) -- (2,1,0) -- (2,2,0);
\draw[yellow!40!black, thick, dashed]
  (0,0,2) -- (1,0,2) -- (1,1,2) -- (2,1,2) -- (2,2,2);

\foreach \t in {0,1,2}{
  \foreach \yy in {0,1,2}{
    \foreach \xx in {0,1}{
      \ifnum\yy=0
        \ifnum\xx=1
        \else
          \draw[link] (\xx,\yy,\t) -- (\xx+1,\yy,\t);
        \fi
      \else
        \draw[link] (\xx,\yy,\t) -- (\xx+1,\yy,\t);
      \fi
    }
  }
}

\foreach \t in {0,1,2}{
  \foreach \xx in {0,1,2}{
    \foreach \yy in {0,1}{
      \ifnum\xx=2
        \ifnum\yy=0
        \else
          \draw[link] (\xx,\yy,\t) -- (\xx,\yy+1,\t);
        \fi
      \else
        \draw[link] (\xx,\yy,\t) -- (\xx,\yy+1,\t);
      \fi
    }
  }
}

\foreach \xx in {0,1,2}{
  \foreach \yy in {0,1,2}{
    \ifnum\xx=0
      \ifnum\yy=0
      \else
        \foreach \t in {0,1}{
          \draw[link] (\xx,\yy,\t) -- (\xx,\yy,\t+1);
        }
      \fi
    \else
      \ifnum\xx=2
        \ifnum\yy=0
        \else
          \ifnum\yy=2
          \else
            \foreach \t in {0,1}{
              \draw[link] (\xx,\yy,\t) -- (\xx,\yy,\t+1);
            }
          \fi
        \fi
      \else
        \foreach \t in {0,1}{
          \draw[link] (\xx,\yy,\t) -- (\xx,\yy,\t+1);
        }
      \fi
    \fi
  }
}

\draw[polyakov, red] (0,0,0) -- (0,0,1);
\draw[polyakov, red] (0,0,1) -- (0,0,2);

\draw[polyakov, blue] (2,2,2) -- (2,2,1);
\draw[polyakov, blue] (2,2,1) -- (2,2,0);

\foreach \t in {0,1,2}{
  \foreach \xx in {0,1,2}{
    \foreach \yy in {0,1,2}{
      \ifnum\xx=2
        \ifnum\yy=0
        \else
          \node[site] at (\xx,\yy,\t) {};
        \fi
      \else
        \node[site] at (\xx,\yy,\t) {};
      \fi
    }
  }
}

\begin{scope}[
  shift={(current bounding box.north east)},
  x={(1cm,0cm)}, y={(0cm,1cm)}, z={(0cm,0cm)}
]
  \node[anchor=north west, align=left, font=\small] at (0.1,-0.1) {
    \textcolor{red}{$\longrightarrow$ Polyakov line ($+q$)}\\[4pt]
    \textcolor{blue}{$\longrightarrow$ Polyakov line ($-q$)}\\[4pt]
    \textcolor{yellow!30!orange}{$\blacksquare$}\ \ Dirac sheet ($x$-links)\\[4pt]
    \textcolor{yellow!60!green}{$\blacksquare$}\ \ Dirac sheet ($y$-links)\\[4pt]
    \textcolor{gray}{$\rule{1cm}{0.6pt}$ gauge link $U_\mu$}
  };
\end{scope}

\end{tikzpicture}
    }
    \caption{Schematic of the lattice setup for two static charges $+q$ and $-q$ at spatial sites $\vec{x}_1$ and $\vec{x}_2$, respectively. The charges trace out Polyakov lines in Euclidean time, which are represented by the red and blue lines. The gray lines represent the gauge links $U_\mu$ on the spatial lattice. Additionally, the Dirac sheet $\Sigma$ for $n^{(q)}_{\mu\nu}(x)$ is displayed in yellow. The stripes are only drawn for better visibility.}
    \label{fig:SC3DwithDirac}
\end{figure}
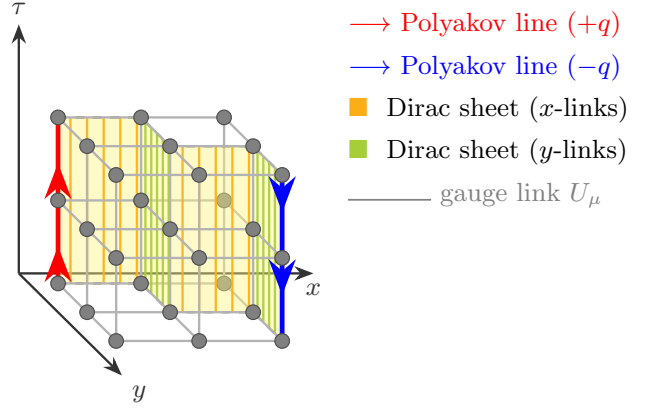

Using
\begin{equation}
\int_{-\pi}^{\pi}\frac{dA}{2\pi}e^{iAk}=\delta_{k,0}\,,
\end{equation}
this leads to
\begin{align}
Z_{q\bar q}
&=
\sum_{\{n_{x,\mu\nu}\}}
\prod_{x,\mu<\nu}
I_{n_{x,\mu\nu}}(\beta)\\
\notag&\quad\times\prod_{x,\mu}
\delta_{
\sum_{\nu}\Delta_\nu n_{\nu\mu}(x)
\,-J_\mu(x),
\,0}\,.
\end{align}
The constraint
\begin{equation}
    \Delta_\nu n_{\nu\mu}(x) \,-J_\mu(x) = 0\,,
\end{equation}
can be fulfilled by choosing
\begin{equation}
\label{eq:nField}
    n_{\mu\nu}(x) = \epsilon_{\mu\nu\rho}\Delta_\rho m(x) + n^{(q)}_{\mu\nu}(x)\,,
\end{equation}
with $\Delta_\mu n_{\mu\nu}^{(q)}(x) = \,J_\nu(x)$.
Here, $m(x)$ is an integer-valued field on the dual lattice sites.
The field $n^{(q)}_{\mu\nu}(x)$ is an integer-valued antisymmetric tensor field. With this combination, the divergence of the lattice curl vanishes and the static integer field fulfills the flux constraint.
If a Dirac sheet $\Sigma$ is chosen such that it has the Polyakov lines of the static charges as its boundary, then $n^{(q)}_{\mu\nu}(x)$ can be chosen to be
\begin{equation}
    n^{(q)}_{\mu\nu}(x) =
\begin{cases}
\pm q\,, & (x,\mu\nu)\in\Sigma,\\
0\,, & \text{otherwise}\,,
\end{cases}
\end{equation}
where the sign of $q$ has to be chosen such that for the specific choice of the Dirac sheet, the divergence at the boundary leads to the correct static charge.
Schematically, this is shown in Fig.~\ref{fig:SC3DwithDirac}.

\subsection{Dualization of the partition function}

To make this step precise, we introduce the Hodge-dual link field on the dual lattice
\begin{equation}
    \tilde n_\rho(X) \coloneq \frac12 \epsilon_{\rho\mu\nu}\, n_{\mu\nu}(x),
\end{equation}
where the dual link $(X,\rho)$ pierces the direct-lattice plaquette $(x,\mu\nu)$.
In (2+1) dimensions, an antisymmetric plaquette field on the direct lattice is thus mapped to an integer-valued link field on the dual lattice. Using Eq.~\eqref{eq:nField},
we obtain
\begin{equation}
    \tilde n_\rho(X)
    = \Delta_\rho m(x) + \tilde n^{(q)}_\rho(X),
    \quad
    \tilde n^{(q)}_\rho(X)\coloneq \frac12 \epsilon_{\rho\mu\nu}\,n^{(q)}_{\mu\nu}(x).
\end{equation}
Hence, the curl part of $n_{\mu\nu}$ becomes a lattice gradient on the dual lattice.

Choose $\gamma$ on the original spatial lattice from $+q$ at $\vec{x}_1$ to $-q$ at $\vec{x}_2$. Extending it in Euclidean time defines the Dirac sheet $\Sigma$. The Hodge-dual background occupies the dual links piercing $\Sigma$; on each spatial slice, these links cross $\gamma$, rather than joining dual sites along a path. Denote it by $n^{(q)}_{ij}=-n^{(\gamma)}_{ij}$, with orientations fixed in Sec.~\ref{sec:theory_electriccharges}.

Similar to Ref.~\cite{Unmuth-Yockey:2018xak}, we can now dualize the partition function.
We find
\begin{equation}
\begin{aligned}
Z_{q\bar q}
&=\sum_{\{ m \}} \prod_t \prod_x I_{m_x(t)-m_x(t+1)}(\beta_s)\\
&\quad\times\prod_{\langle ij \rangle} I_{m_i(t)-m_j(t)-n^{(q)}_{ij}}(\beta_t)\\
&=\sum_{\{ m \}} \prod_{t=1}^{N_t} K\bigl(m(t),m(t+1)\bigr) V\bigl(m(t)\bigr)\,,
\end{aligned}
\end{equation}
with
\begin{align}
K\bigl(m(t),m(t+1)\bigr)
&= \prod_x I_{m_x(t)-m_x(t+1)}(\beta_s)\,,\\
V\bigl(m(t)\bigr)
&= \prod_{\langle ij \rangle} I_{m_i(t)-m_j(t)-n^{(q)}_{ij}}(\beta_t)\,.
\end{align}

\subsection{Obtaining the Hamiltonian}
From this, the transfer matrix can be identified as
\begin{equation}
\langle m(t+1)|\hat{T}|m(t)\rangle
= K\bigl(m(t),m(t+1)\bigr) V\bigl(m(t)\bigr)\,.
\end{equation}
If one now introduces $\hat{m}_x|m\rangle = m_x |m\rangle$, with $m_x \in \mathbb{Z}$ and the ladder operators
\begin{equation}
    e^{\pm i \hat{\phi}_x} |m_x\rangle = |m_x \pm 1\rangle\,,
\end{equation}
the matrix elements can be expressed in the integer basis.
Using
\begin{equation}
    e^{\beta \cos\hat{\phi}}
    =
    \sum_{n\in\mathbb Z} I_n(\beta)\,e^{in\hat{\phi}}\,,
\end{equation}
one obtains
\begin{equation}
\begin{aligned}
    \langle l | e^{\beta \cos\hat{\phi}} | m \rangle
    &= \sum_{n\in\mathbb Z} I_n(\beta)\,\langle l|e^{in\hat{\phi}}|m\rangle \\
    &= \sum_{n\in\mathbb Z} I_n(\beta)\,\delta_{l,m+n} \\
    &= I_{l-m}(\beta)\,.
\end{aligned}
\end{equation}
Hence, one finds
\begin{align}
K\bigl(m(t),m(t+1)\bigr)
&= \langle m(t+1) | \prod_x e^{\beta_s \cos\hat{\phi}_x} | m(t) \rangle\,.
\end{align}
As $V\bigl(m(t)\bigr)$ is diagonal in this basis, one obtains
\begin{align}
\hat{V}|m\rangle &= V\left(m\right)|m\rangle\\
V\left(m\right) &= \prod_{\langle ij \rangle} I_{m_i - m_j - n^{(q)}_{ij}}(\beta_t)\,.
\end{align}
In the following, we assume two different couplings for the spatial and time directions on the direct lattice.
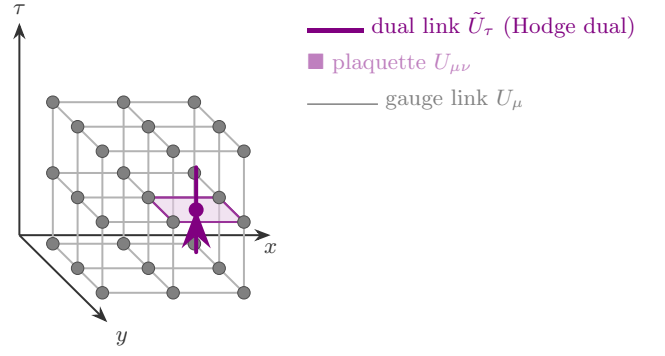
\begin{figure}[!t]
    \centering
    \resizebox{\linewidth}{!}{%
    \begin{tikzpicture}[
  scale=1.0,
  x={(1cm,0cm)},
  y={({cos(60)*0.7cm},{sin(60)*-0.4cm})},
  z={(0cm,1.0cm)},
  site/.style={circle, fill=black!50, draw=black!70, inner sep=0pt, minimum size=5pt},
  link/.style={thick, gray!60},
  duallink/.style={ultra thick, violet, line cap=round,
    postaction={decorate, decoration={
      markings,
      mark=at position 0.5 with {\arrow[scale=1.5]{Stealth}}
    }}
  },
  plaq/.style={fill=violet!20, fill opacity=0.55, draw=violet!70, thick},
]

\def\tmin{0}
\def\tmax{2}

\draw[-Stealth, thick, black!80] (-0.35,-0.35,0) -- (3.2,-0.35,0)
  node[below, font=\small] {$x$};
\draw[-Stealth, thick, black!80] (-0.35,-0.35,0) -- (-0.35,3.2,0)
  node[below right, font=\small] {$y$};
\draw[-Stealth, thick, black!80] (-0.35,-0.35,0) -- (-0.35,-0.35,3.0)
  node[above, font=\small] {$\tau$};

\foreach \t in {0,1,2}{
  \foreach \yy in {0,1,2}{
    \foreach \xx in {0,1}{
      \draw[link] (\xx,\yy,\t) -- (\xx+1,\yy,\t);
    }
  }
  \foreach \xx in {0,1,2}{
    \foreach \yy in {0,1}{
      \draw[link] (\xx,\yy,\t) -- (\xx,\yy+1,\t);
    }
  }
}

\foreach \xx in {0,1,2}{
  \foreach \yy in {0,1,2}{
    \foreach \t in {0,1}{
      \draw[link] (\xx,\yy,\t) -- (\xx,\yy,\t+1);
    }
  }
}

\fill[plaq]
  (1,1,1) -- (2,1,1) -- (2,2,1) -- (1,2,1) -- cycle;
\draw[violet!80, thick]
  (1,1,1) -- (2,1,1) -- (2,2,1) -- (1,2,1) -- cycle;

\foreach \t in {0,1,2}{
  \foreach \xx in {0,1,2}{
    \foreach \yy in {0,1,2}{
      \node[site] at (\xx,\yy,\t) {};
    }
  }
}

\draw[duallink, violet]
  (1.5, 1.5, 0.4) -- (1.5, 1.5, 1.6);

\node[circle, fill=violet, inner sep=0pt, minimum size=6pt] at (1.5,1.5,1) {};

\begin{scope}[
  shift={(current bounding box.north east)},
  x={(1cm,0cm)}, y={(0cm,1cm)}, z={(0cm,0cm)}
]
  \node[anchor=north west, align=left, font=\small] at (0.2,-0.1) {
    \textcolor{violet}{$\rule{0.8cm}{2pt}$ dual link $\tilde{U}_\tau$ (Hodge dual)}\\[4pt]
    \textcolor{violet!50}{$\blacksquare$ plaquette $U_{\mu\nu}$}\\[4pt]
    \textcolor{gray}{$\rule{1cm}{0.6pt}$ gauge link $U_\mu$}
  };
\end{scope}

\end{tikzpicture}
    }
    \caption{Schematic of the Hodge dual of a single plaquette. The plaquette becomes a direct link on the dual lattice, thus now coupling two positions in the time direction.}
    \label{fig:Dual}
\end{figure}
Note that because links pierce direct plaquettes orthogonally on the dual lattice, the spatial coupling appears in the time step and vice versa as shown in Fig.~\ref{fig:Dual}.
The Hamiltonian can now be obtained by taking the continuous-time limit $\beta_t \to \infty$.
Assuming $|j| \ll \alpha$ the Bessel functions can be expanded as
\begin{equation}
I_j(\alpha) \approx \frac{e^\alpha}{\sqrt{2\pi\alpha}} \exp\!\left(-\frac{j^2}{2\alpha}\right).
\end{equation}
This leads to
\begin{equation}
\begin{aligned}
V\left(m\right) &= \prod_{\langle ij \rangle} I_{m_i - m_j - n^{(q)}_{ij}}(\beta_t)\\
-\log\!V\left(m\right)
&\approx \text{const.} + \frac{1}{2\beta_t} \sum_{\langle ij \rangle} (m_i - m_j - n^{(q)}_{ij})^2\,.
\end{aligned}
\end{equation}
This expansion is valid in the large-$\beta_t$ (continuous-time) limit, where electric field fluctuations are small compared to $\beta_t$.
For the second part one finds
\begin{equation}
\begin{aligned}
\hat{K}
&= \prod_x e^{\beta_s \cos\hat{\phi}_x}\\
-\log\!\hat{K}
&= -\beta_s \sum_x \cos\hat{\phi}_x\\
&= -\frac{\beta_s}{2} \sum_x (\hat{P}_x + \hat{P}^\dagger_x)\,.
\end{aligned}
\end{equation}
Combining these two parts and taking the continuous-time limit then yields the Hamiltonian.
As in Ref.~\cite{Unmuth-Yockey:2018xak}, the couplings can be defined as
\begin{equation}
    U \coloneq \frac{1}{\beta_t a},\qquad X \coloneq \frac{\beta_s}{a},
\end{equation}
where $a$ is the lattice spacing in the time direction. Note that we consider the limits $\beta_s\to0$, $\beta_t\to\infty$, and $a\to 0$, such that $U=\text{const.}$ and $X=\text{const.}$
This leads to the final form of the dual Hamiltonian with static charges
\begin{equation}\label{eq:dual_hamiltonian_sc_UX}
    \hat{H}_{\text{dual}}=\frac{U}{2}\sum_{\langle ij \rangle} (\hat{L}^Z_i - \hat{L}^Z_j - n^{(q)}_{ij})^2 - \frac{X}{2} \sum_i (\hat{P}_i + \hat{P}^{\dagger}_i)\,,
\end{equation}
where $\hat{L}^Z_i |m_i\rangle = m_i |m_i\rangle$.

Using the convention stated below Eq.~\eqref{eq:link_hamiltonian}, the magnetic term contains the additive constant $N_{\mathrm{plaq}}/g^2$. Dropping this constant leaves $-(2g^2)^{-1}\sum_i(\hat P_i+\hat P_i^\dagger)$, which gives $X=1/g^2$ in Eq.~\eqref{eq:dual_hamiltonian_sc_UX}, while comparison of the electric terms gives $U=g^2$. Equation~\eqref{eq:dual_hamiltonian_sc_UX} therefore agrees with Eq.~\eqref{eq:dual_hamiltonian_sc}, with the sign difference absorbed into the relation between $n^{(q)}_{ij}$ and the main-text background field $n^{(\gamma)}_{ij}$. For the same link orientation, the path background used in the main text is $n^{(\gamma)}_{ij}=-n^{(q)}_{ij}$.

  \section{Uncontrolled continuum extrapolation on small volumes}\label{app:uncontrolled}

\begin{figure}[!htbp]
    \centering
    \includegraphics[width=\linewidth]{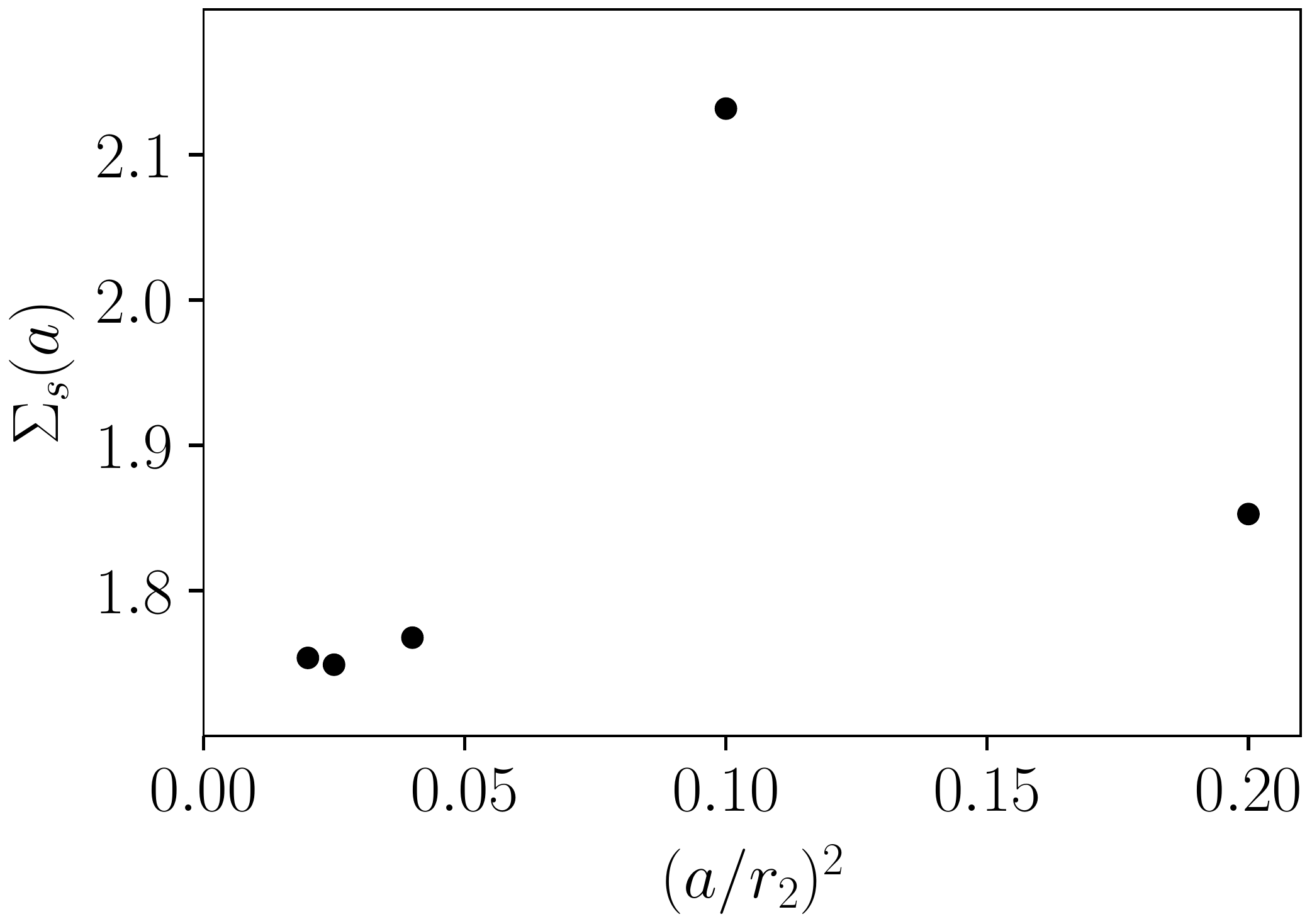}
    \caption{Uncontrolled continuum extrapolation of the lattice step-scaling function $\Sigma_s\left(a\right)$ as a function of $\left(a/r_2\right)^2$, performed on small volumes from $3\times 3$ up to $7\times 7$ at truncation $\Lambda=1$.}
    \label{fig:CLSS_uncontrolled}
\end{figure}

As a negative control for the finite-volume scheme of Sec.~\ref{sec:theory_FVS}, we show in Fig.~\ref{fig:CLSS_uncontrolled} what the lattice step-scaling function looks like when the ratio $r_1/N_x$ is not held fixed along the continuum sequence. We vary the lattice size from $3\times 3$ up to $7\times 7$ without matching the physical reference distance across discretizations, and plot $\Sigma_s$ as a function of $(a/r_2)^2$.

The resulting trajectory is irregular and non-monotonic: a varying finite-volume contribution is folded into what would otherwise be the cutoff dependence, masking the continuum trend. This benchmark motivates the fixed-ratio construction adopted in Sec.~\ref{sec:theory_FVS}.

  \section{Path dependence at finite local-basis truncation}
\label{app:path_deformation}

This appendix specifies the background path used in the production calculations, shows why the path is an integer-field relabeling in the untruncated theory, and measures the residual path dependence caused by the finite local basis.

\subsection{Integer-field relabeling}

Let $m_i\in\mathbb{Z}$ be the plaquette-flux variable at dual site $i$. On a link $\langle ij\rangle$, the electric term depends on
\begin{equation}
    \ell_{ij}^{(\gamma)}=m_i-m_j+n_{ij}^{(\gamma)}\,.
\end{equation}
For two paths $\gamma$ and $\gamma'$ with the same endpoints, let $k_i\in\mathbb{Z}$ be a fixed integer assigned to each plaquette (equivalently, each dual site); for a simple deformation, $k_i=\pm1$ on the plaquettes enclosed by $\gamma'-\gamma$ and zero elsewhere. Because the lattice is open and simply connected, $\gamma'-\gamma$ is the boundary of this plaquette field, so, with our orientation convention,
\begin{equation}
    n_{ij}^{(\gamma')}=n_{ij}^{(\gamma)}-(k_i-k_j)\,.
\end{equation}
Relabeling every integer variable by $m_i'=m_i+k_i$ then gives, link by link,
\begin{align}
    m_i'-m_j'+n_{ij}^{(\gamma')}
    &=m_i+k_i-m_j-k_j+n_{ij}^{(\gamma)}-(k_i-k_j)\notag\\
    &=m_i-m_j+n_{ij}^{(\gamma)}\,.
\end{align}
The map $\{m_i\}\mapsto\{m_i'\}$ is a bijection of the untruncated integer configuration space, and the unit translations commute with the magnetic shift operators. Thus the two Hamiltonians are unitarily equivalent and have identical spectra. This is the Hamiltonian counterpart of changing the Dirac sheet in the transfer-matrix construction of Appendix~\ref{sec:app}.

At finite $\Lambda$, the retained single-plaquette subspace is not invariant under $m_i\mapsto m_i+k_i$. Equivalently, the finite-basis projector does not commute with the integer translation. The preceding unitary equivalence is then broken by the projection, so a residual dependence on $\gamma$ is expected.

\subsection{Numerical comparison at \texorpdfstring{$\Lambda=1$}{Lambda=1}}

\begin{figure*}[t]
    \centering
    \includegraphics[width=\textwidth]{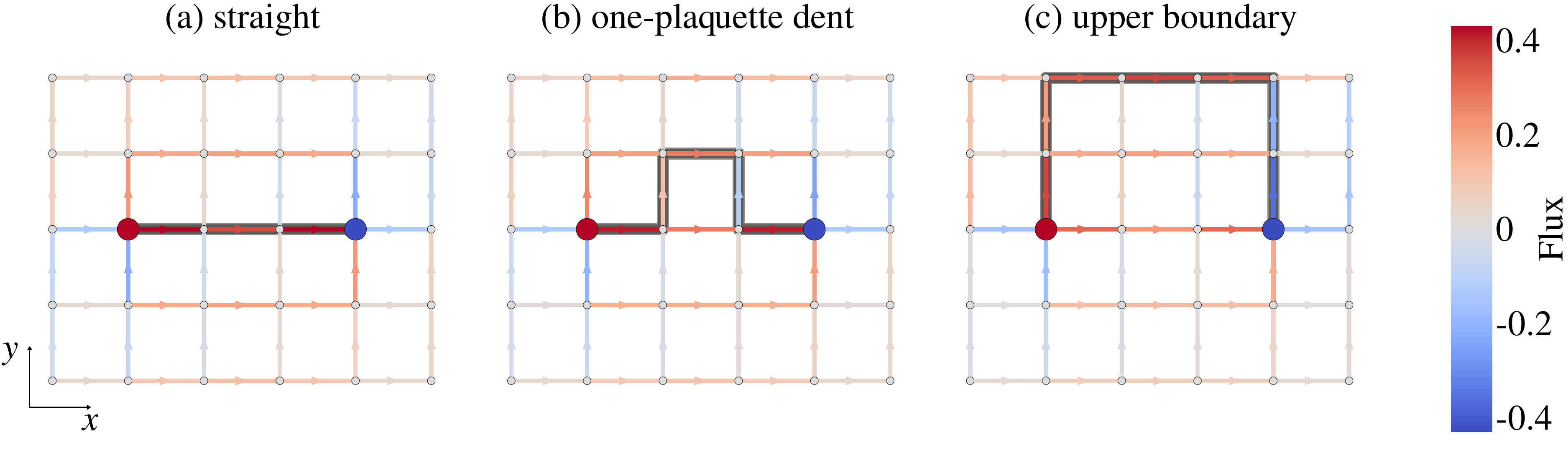}
    \caption{Oriented electric-flux expectation values for the straight path, the one-plaquette dent, and the upper-boundary path on the $6\times5$ lattice at $g^2=0.5$, $\Lambda=1$, and final SVD cutoff $10^{-15}$. All panels use the same color scale; arrows point in the positive link orientation, and the sign is encoded by color. The translucent grey band marks the imposed integer background path, while the red and blue circles mark the $+1$ and $-1$ charges. The measured flux obeys Gauss' law in every panel, but its transverse profile shifts toward the chosen path.}
    \label{fig:path_deformation_flux}
\end{figure*}

\begin{table*}[!t]

\begin{ruledtabular}
\begin{tabular}{cccrrrrr}
$N_y$ & $g^2$ & sector & $V_\gamma$ & $\Delta_\gamma E$ & relative (\%) & $|\mathrm{Var}(H)|$ & $\chi_{\max}$ \\
\hline
3 & 0.50 & $\gamma_0$             & 0.345514 & 0 & 0 & $2.1\times10^{-13}$ & 34 \\
3 & 0.50 & $\gamma_1$             & 0.370101 & 0.024587 & 7.12 & $3.4\times10^{-13}$ & 32 \\
3 & 0.50 & $\gamma_{\mathrm b}$   & 0.418159 & 0.072645 & 21.03 & $1.4\times10^{-13}$ & 32 \\
3 & 0.25 & $\gamma_0$             & 0.166494 & 0 & 0 & $5.8\times10^{-13}$ & 34 \\
3 & 0.25 & $\gamma_1$             & 0.174175 & 0.007681 & 4.61 & $5.4\times10^{-13}$ & 33 \\
3 & 0.25 & $\gamma_{\mathrm b}$   & 0.189828 & 0.023334 & 14.02 & $4.7\times10^{-13}$ & 33 \\
5 & 0.50 & $\gamma_0$             & 0.299980 & 0 & 0 & $3.9\times10^{-12}$ & 300 \\
5 & 0.50 & $\gamma_1$             & 0.321193 & 0.021212 & 7.07 & $4.2\times10^{-12}$ & 286 \\
5 & 0.50 & $\gamma_{\mathrm b}$   & 0.473996 & 0.174016 & 58.01 & $1.7\times10^{-12}$ & 271 \\
5 & 0.25 & $\gamma_0$             & 0.143558 & 0 & 0 & $1.3\times10^{-12}$ & 301 \\
5 & 0.25 & $\gamma_1$             & 0.151422 & 0.007865 & 5.48 & $4.3\times10^{-12}$ & 294 \\
5 & 0.25 & $\gamma_{\mathrm b}$   & 0.204170 & 0.060612 & 42.22 & $4.8\times10^{-12}$ & 283 \\
\end{tabular}
\end{ruledtabular}

\begin{ruledtabular}
\begin{tabular}{crrrrrrr}
\multicolumn{8}{c}{$\Lambda$ scan with $N_x=6$, $N_y=5$, $g^2=0.25$, $r/a=3$, final SVD cutoff $10^{-9}$} \\
$\Lambda$ & $V_{\mathrm{straight}}$ & $\Delta_{\gamma_1} E$ & relative (\%) & $\Delta_{\gamma_{\mathrm b}} E$ & relative (\%) & max $|\mathrm{Var}(H)|$ & $\chi_{\max}$ \\
\hline
1 & 0.143558 & $7.865\times10^{-3}$ & 5.48 & $6.061\times10^{-2}$ & 42.2 & $3.21\times10^{-6}$ & 61 \\
2 & 0.131760 & $9.097\times10^{-4}$ & 0.690 & $7.721\times10^{-3}$ & 5.86 & $4.01\times10^{-6}$ & 92 \\
3 & 0.130690 & $6.820\times10^{-5}$ & 0.0522 & $5.467\times10^{-4}$ & 0.418 & $4.26\times10^{-6}$ & 99 \\
\end{tabular}
\end{ruledtabular}
\caption{Path-deformation results with the first block at $\Lambda=1$ and final SVD cutoff $10^{-15}$, and a $\Lambda=1,2,3$ scan at final SVD cutoff $10^{-9}$ in the second block. Energies are in lattice units, $V_\gamma=E_\gamma-E_{\mathrm{vac}}$, and $\Delta_\gamma E=E_\gamma-E_{\gamma_0}$. The relative shift is normalized by the straight-path potential. The tables report the absolute energy variance and maximum MPS bond dimension.}
\label{tab:path_deformation}
\end{table*}

We compare the three paths at fixed charge positions on $6\times3$ and $6\times5$ lattices. The charges are at $(1,y_0)$ and $(4,y_0)$, where $y_0=(N_y-1)/2$ and the paths are
\begin{align}
\gamma_0={}&(1,y_0)\to(2,y_0)\to(3,y_0)\to(4,y_0)\,,\notag\\
\gamma_1={}&(1,y_0)\to(2,y_0)\to(2,y_0+1)\to(3,y_0+1)\notag\\
&\to(3,y_0)\to(4,y_0)\,,\notag\\
\gamma_{\mathrm b}={}&(1,y_0)\to\cdots\to(1,N_y-1)\to(2,N_y-1)\notag\\
&\to(3,N_y-1)\to(4,N_y-1)\to\cdots\to(4,y_0)\,.
\end{align}
$\gamma_0$ is the straight horizontal path used in the production simulations; $\gamma_1$ forms a one-plaquette ``dent'' relative to $\gamma_0$, while $\gamma_{\mathrm b}$ runs along the upper boundary; lower deformations are related by reflection. For each $N_y\in\{3,5\}$ and $g^2\in\{0.5,0.25\}$, we hold the open boundary conditions, charge positions, $\Lambda=1$ basis, DMRG schedule, and initialization fixed, changing only the ordered path. We use one vacuum energy for all three charged sectors. These results use final SVD cutoff $10^{-15}$.

Defining
\begin{equation}
    V_\gamma=E_\gamma-E_{\mathrm{vac}},\qquad
    \Delta_\gamma E=V_\gamma-V_{\gamma_0}=E_\gamma-E_{\gamma_0}\,,
\end{equation}
the complete results are given in the first block of Table~\ref{tab:path_deformation}. Every tested deformation raises the optimized charged-sector energy relative to the straight path.

At $\Lambda=1$, the one-plaquette residual is $4.6\%$--$7.1\%$ of the straight-path potential and is stable across $N_y=3,5$; the upper-boundary residual is larger, $14\%$--$58\%$. For $N_y=3$, $\gamma_1$ and $\gamma_{\mathrm b}$ both have five links, yet the boundary residual is about three times larger because the paths enclose respectively one and three plaquettes. For $N_y=5$, the boundary path encloses six plaquettes and gives a residual $7.7$--$8.2$ times the one-plaquette value, indicating additional boundary effects.

Because $\gamma$ specifies only the background integer field and does not constrain the physical flux, the residual is a finite-cutoff relabeling artifact, not physical energy of a longer string.

In every charged run, the measured horizontal flux through the cut midway between the charges sums to one, as required by Gauss' law. Figure~\ref{fig:path_deformation_flux} shows the $N_y=5$, $g^2=0.5$ case.

The separation dependence of these residuals is tested below; it determines whether they cancel in a force.

\subsection{Scan in the local-basis cutoff}

We repeat the fixed-position comparison for $\Lambda=1,2,3$ on the $N_x=6$, $N_y=5$ lattice at $g^2=0.25$ and $r/a=3$, using the same straight, one-plaquette dent, and upper-boundary paths. All three results use final SVD cutoff $10^{-9}$. Only the local basis is changed. The $\Lambda=1$ values agree with results at cutoff $10^{-15}$, for which the largest difference in a path residual is $6.84\times10^{-9}$. For each $\Lambda$, the vacuum energy is common to the three charged sectors; the maximum variances and bond dimensions are shown in the table.

The energy for each fixed path decreases toward the common untruncated value as $\Lambda$ increases. At $\Lambda=3$, the $\gamma_1$ and $\gamma_{\mathrm b}$ residuals are respectively $0.00867$ and $0.00902$ of their $\Lambda=1$ values, corresponding to approximately $99.1\%$ reductions. Thus the scan supports restoration of path independence and we can identify the positive finite-$\Lambda$ path excess as due to the Hilbert space truncation. These three $\Lambda$ values do not determine an asymptotic law or independently prove the exact $\Lambda\to\infty$ limit, which follows from the integer-field relabeling argument above. Because increasing $\Lambda$ lowers each fixed-path variational energy toward the common untruncated value, while the tested deformations raise the $\Lambda=1$ energy relative to the straight path, the straight path gives the lowest variational upper bound, and is therefore the preferred finite-$\Lambda$ representation, among the paths tested.

\subsection{Separation dependence}

At each separation we compare paths with identical charge positions on an $N_x=8$ lattice at $N_y=3,5$, $g^2=0.25,0.5$, and $\Lambda=1,2,3$, using the straight, one-plaquette dent, and upper-boundary paths at $r/a=3,5$. All runs use final SVD cutoff $10^{-9}$. With one common vacuum energy, define
\begin{equation}
    \begin{aligned}
        S_0&=\frac{V_{\gamma_0}(5a)-V_{\gamma_0}(3a)}{2},\\
        \delta S_\gamma&=\frac{\Delta_\gamma E(5a)-\Delta_\gamma E(3a)}{2},\\
        R_\gamma&\equiv\frac{\delta S_\gamma}{S_0}.
    \end{aligned}
\end{equation}
Note that the excess link count is independent of $r$, as it is $+2$ for the dent, $+2$ for the boundary path at $N_y=3$, and $+4$ for the boundary path at $N_y=5$. Thus a nonzero $\delta S_\gamma$ is not a constant path-length offset (Table~\ref{tab:path_separation}).

\begin{table}[!t]

\begin{ruledtabular}
\begin{tabular}{cccrrr}
$N_y$ & $g^2$ & path & $\Lambda=1$ & $\Lambda=2$ & $\Lambda=3$ \\
\hline
3 & 0.25 & dent     & $-0.957\%$ & $-0.0898\%$ & $-0.00508\%$ \\
3 & 0.25 & boundary & $+11.6\%$ & $+1.13\%$ & $+0.0586\%$ \\
3 & 0.50 & dent     & $-1.39\%$ & $-0.0448\%$ & $-0.00022\%$ \\
3 & 0.50 & boundary & $+16.5\%$ & $+0.553\%$ & $+0.00224\%$ \\
5 & 0.25 & dent     & $-2.45\%$ & $-0.397\%$ & $-0.0323\%$ \\
5 & 0.25 & boundary & $+33.2\%$ & $+5.86\%$ & $+0.405\%$ \\
5 & 0.50 & dent     & $-2.86\%$ & $-0.161\%$ & $-0.00116\%$ \\
5 & 0.50 & boundary & $+42.4\%$ & $+3.07\%$ & $+0.0173\%$ \\
\end{tabular}
\end{ruledtabular}
\caption{Signed separation shifts $R_\gamma\times100\%$; positive (negative) $R_\gamma$ means that the deformation raises (lowers) the secant slope relative to $\gamma_0$.}
\label{tab:path_separation}
\end{table}

The residuals of the path with a one-plaquette dent decrease with separation, $\Delta_\gamma E(5a)/\Delta_\gamma E(3a)=0.755$--$0.894$, while boundary residuals increase, with ratios $1.345$--$1.449$. Increasing $\Lambda$ suppresses the artifact, but at $\Lambda=1$ it does not generally cancel in a force.

  \section{Potential-fit-range sensitivity of the logarithmic coefficient}
\label{app:potential_fit_range}

Using the continuum-sequence data, we fit
\begin{equation}
    aV(r/a)=\alpha+\sigma(r/a)+\delta\log(r/a)
\end{equation}
over the production range $r/a=1,\ldots,N_x-3$ and the shortened range $r/a=2,\ldots,N_x-3$. At $N_x=7$, the shortened range has exactly three points for three fit parameters and hence zero residual degrees of freedom. We omit $N_x=6$ because removing $r/a=1$ would leave only two points for three parameters.

We apply this scan to the $(\Lambda,N_y)=(1,3)$, $(2,3)$, and $(2,5)$ sequences. The resulting dimensionless coefficients and their changes are shown in Table~\ref{tab:potential_fit_range}. Omitting the first point lowers the positive $\delta$ by $47.9$--$61.2\%$ for $(1,3)$, $49.7$--$61.6\%$ for $(2,3)$, and $42.2$--$47.6\%$ for $(2,5)$.

\begin{table}[!htbp]
\begin{ruledtabular}
\small
\setlength{\tabcolsep}{1pt}
\begin{tabular}{c c c c c c c}
$\Lambda$ & $N_y$ & $N_x$ & $g^2_{\rm tuned}$ & $\delta_{r\geq1}$ & $\delta_{r\geq2}$ & reduction (\%) \\
\hline
1 & 3 & 7  & 0.384650 & 0.019635 & 0.010232 & 47.9 \\
1 & 3 & 8  & 0.302176 & 0.013893 & 0.006755 & 51.4 \\
1 & 3 & 9  & 0.242630 & 0.010069 & 0.004572 & 54.6 \\
1 & 3 & 10 & 0.198636 & 0.007481 & 0.003200 & 57.2 \\
1 & 3 & 11 & 0.165391 & 0.005691 & 0.002311 & 59.4 \\
1 & 3 & 12 & 0.139735 & 0.004422 & 0.001717 & 61.2 \\
2 & 3 & 7  & 0.409698 & 0.023086 & 0.011605 & 49.7 \\
2 & 3 & 8  & 0.319361 & 0.015933 & 0.007583 & 52.4 \\
2 & 3 & 9  & 0.254624 & 0.011332 & 0.005064 & 55.3 \\
2 & 3 & 10 & 0.207691 & 0.008336 & 0.003517 & 57.8 \\
2 & 3 & 11 & 0.172471 & 0.006295 & 0.002526 & 59.9 \\
2 & 3 & 12 & 0.145418 & 0.004865 & 0.001868 & 61.6 \\
2 & 5 & 7  & 0.548080 & 0.049802 & 0.027736 & 44.3 \\
2 & 5 & 8  & 0.468054 & 0.040603 & 0.023453 & 42.2 \\
2 & 5 & 9  & 0.387745 & 0.031304 & 0.017982 & 42.6 \\
2 & 5 & 10 & 0.319371 & 0.023772 & 0.013287 & 44.1 \\
2 & 5 & 11 & 0.267622 & 0.018484 & 0.009999 & 45.9 \\
2 & 5 & 12 & 0.228477 & 0.014765 & 0.007735 & 47.6 \\
\end{tabular}
\end{ruledtabular}
\caption{Sensitivity of the fitted logarithmic coefficient to the potential-fit range. The table lists the sequence parameters, tuned couplings, and coefficients from the $r/a\geq1$ and $r/a\geq2$ potential-fit ranges. The final column gives the relative reduction in percent.}
\label{tab:potential_fit_range}
\end{table}

This scan changes only the potential-fit range used to extract $\delta$; it does not propagate the change through the force, tuning, or step-scaling analyses.

For completeness, the same scan was applied to the preliminary $\Lambda=2$
$N_x=16,20$ extension (Table~\ref{tab:potential_fit_range_extension}). The relative reductions are $66.2$--$68.9\%$ for
$N_y=3$ and $52.8$--$56.1\%$ for $N_y=5$; these points are not included in
the $N_x\leq12$ scan in Table~\ref{tab:potential_fit_range}.

\begin{table}[!htbp]
\begin{ruledtabular}
\small
\setlength{\tabcolsep}{2pt}
\begin{tabular}{c c c c c c}
$N_y$ & $N_x$ & $g^2_{\rm tuned}$ & $\delta_{r\geq1}$ & $\delta_{r\geq2}$ & reduction (\%) \\
\hline
3 & 16 & 0.082392 & 0.002072 & 0.000700 & 66.2 \\
3 & 20 & 0.052894 & 0.001071 & 0.000333 & 68.9 \\
5 & 16 & 0.133180 & 0.006805 & 0.003214 & 52.8 \\
5 & 20 & 0.086423 & 0.003647 & 0.001602 & 56.1 \\
\end{tabular}
\end{ruledtabular}
\caption{Preliminary $\Lambda=2$, $N_x=16,20$ extension fit-window scan.
The final column gives the relative reduction on omitting $r/a=1$.}
\label{tab:potential_fit_range_extension}
\end{table}

  {
\setlength{\parskip}{0.2mm}
\setlength{\abovedisplayskip}{2pt plus 1pt minus 1pt}
\setlength{\belowdisplayskip}{2pt plus 1pt minus 1pt}
\setlength{\abovedisplayshortskip}{1pt plus 1pt minus 1pt}
\setlength{\belowdisplayshortskip}{1pt plus 1pt minus 1pt}
\setlength{\intextsep}{8pt plus 2pt minus 2pt}
\setlength{\textfloatsep}{8pt plus 2pt minus 2pt}
\setlength{\dbltextfloatsep}{4pt plus 2pt minus 2pt}
\captionsetup{skip=2pt}
\section{Weak-coupling behavior and basis dependence}\label{app:weak_lambda}
\label{app:continuum_extension}

\begin{figure*}[!tp]
    \centering
    \includegraphics[width=\textwidth]{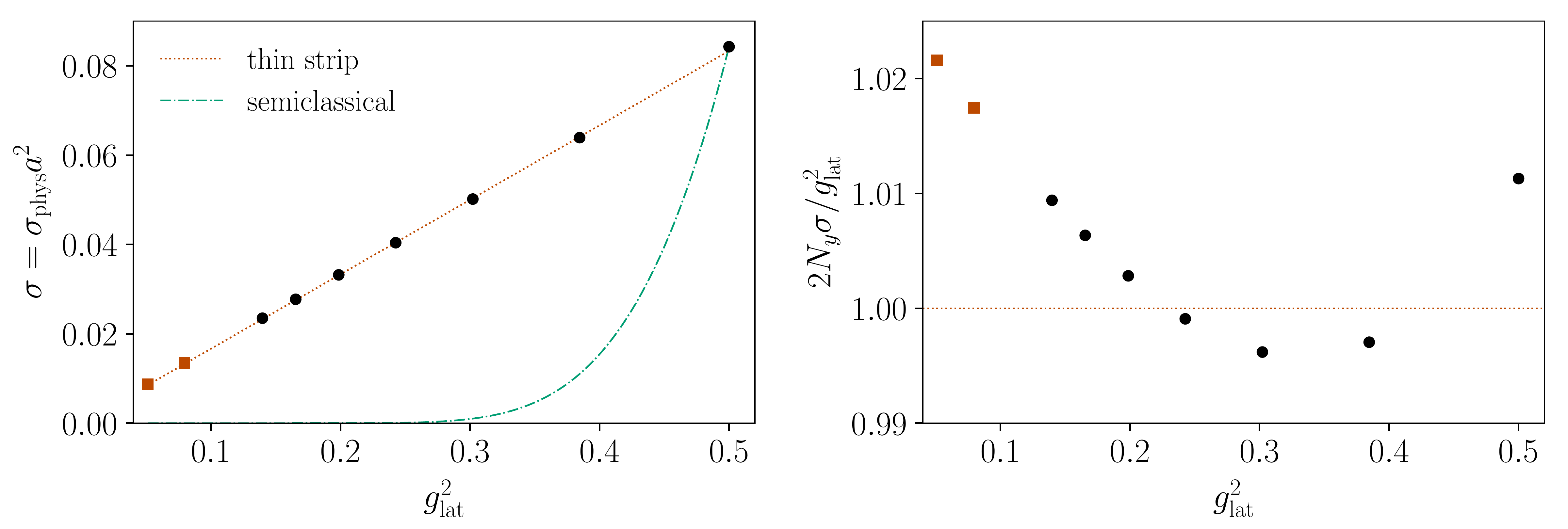}
    \caption{Fitted string coefficient at $N_y=3$, $\Lambda=1$, including
    $N_x=16,20$ (orange squares). Left: $\sigma$, the thin-strip scale
    $g_{\rm lat}^2/(2N_y)$ (orange dotted), and the semiclassical reference
    (green dash-dotted), normalized at the coarsest point.
    Right: the ratio to the strip scale.}
    \label{fig:weak_sigma_scaling}
\end{figure*}

At weak coupling, monopoles generate an exponentially small string tension
in compact $(2+1)$-dimensional $\Uone$ gauge theory,
\begin{equation}
  \sigma\sim C(g_{\rm lat}^2)\exp[-c/g_{\rm lat}^2],
\end{equation}
where $C$ is a coupling-dependent prefactor~\cite{Polyakov:1975rs,Banks:1977cc,Gopfert:1981yf}.

A finite approximation of the gauge group can change this behavior.
The zero-temperature three-dimensional vector $\mathbb{Z}_N$ gauge theory,
for example, has a weak-coupling phase with vanishing string tension.
Its transition scale satisfies
$\beta_c\propto[1-\cos(2\pi/N)]^{-1}\sim N^2$, hence
$g_c^2\sim N^{-2}$~\cite{Borisenko:2013qza}.
Increasing $N$ moves the transition toward weaker coupling, but at fixed
$N$ lowering $g^2$ eventually enters the nonconfining phase.

Our single-plaquette truncation is not a $\mathbb{Z}_N$ approximation.
This comparison nevertheless motivates checking its weak-coupling behavior
as $\Lambda$ increases. We follow $\sigma$ and $\delta$ along the tuned
sequences, compare the string coefficient with the semiclassical prediction
and the finite-width contribution, and test the basis dependence of both
the weak-coupling observable and the step-scaling extrapolation.

\subsection{String coefficient and transverse width}

Let $R=r/a$. With the closed-flux spatial boundaries used here, Gauss' law
fixes the total longitudinal flux through each of the $R$ transverse cuts
between unit charges:
\begin{equation}
\sum_{y=0}^{N_y-1}\langle\hat E_x(c,y)\rangle=1,
\qquad c=1,\ldots,R.
\end{equation}

Positivity of the variance and Cauchy--Schwarz give, for each cut,
\begin{align}
\sum_y\langle\hat E_x(c,y)^2\rangle
&\geq \sum_y\langle\hat E_x(c,y)\rangle^2 \\
&\geq \frac{1}{N_y}\left(\sum_y
\langle\hat E_x(c,y)\rangle\right)^2
=\frac{1}{N_y}.
\end{align}
Therefore the electric term between the charges obeys
\begin{equation}
\left\langle\hat H_{E,\mathrm{between}}\right\rangle
\geq R\frac{g_{\rm lat}^2}{2N_y}.
\end{equation}
\begin{figure*}[!tp]
    \centering
    \includegraphics[width=\textwidth]{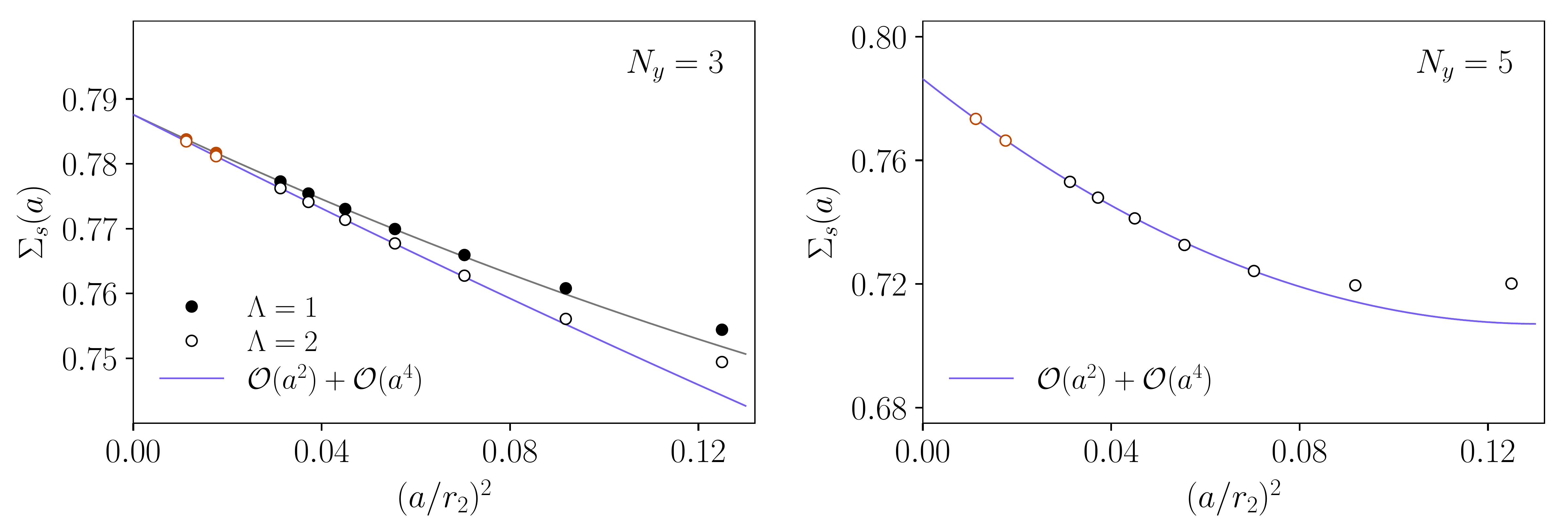}
    \caption{Step-scaling at $\Lambda=1$ (filled circles, gray fit) and
    $\Lambda=2$ (open circles, purple fit). Only the $f_3$ fit, with
    $a^2$ and $a^4$ corrections, is shown, using $N_x=10,11,12,16,20$.
    Orange points are $N_x=16,20$; only $\Lambda=2$ is available at $N_y=5$.
    Curves are continued outside the fit range. The $\Lambda=2$ extension
    points are preliminary.}
    \label{fig:lambda2_continuum}\label{fig:continuum_extension}
    \label{fig:continuum_lambda2_extension}
\end{figure*}

The electric scale per unit length is thus $g_{\rm lat}^2/(2N_y)$.

The closed-flux boundaries prevent transverse spreading beyond the strip. In
the Maxwell approximation, a strip of width $N_y$ therefore supports a
linear potential at separations large compared with its width, with
coefficient $g_{\rm lat}^2/(2N_y)$, even without monopoles. This finite-width
contribution can mask a bulk monopole tension that is exponentially smaller at
weak coupling. A corresponding weak-coupling finite-size scale is reported in
Ref.~\cite[Eq.~(28)]{Loan:2002ej}.

Given that the potential includes both electric and magnetic contributions, the electric-energy bound does not apply directly to the fitted slope $\sigma$.
Subtracting $g_{\rm lat}^2/(2N_y)$ therefore does not isolate a monopole
contribution. In Eq.~\eqref{eq:potential_fit},
$\sigma=\sigma_{\rm phys}a^2$. Table~\ref{tab:weak_sigma_delta} uses the
production fit range $r/a=1,\ldots,N_x-3$.

\begin{table}[!htbp]
\centering

\begin{tabular}{ccccc}
$N_x$ & $g_{\rm lat}^2$ & $\sigma=\sigma_{\rm phys}a^2$ & $\delta$ &
$2N_y\sigma/g_{\rm lat}^2$ \\
\hline
6  & 0.500000 & 0.084274666 & 0.028360583 & 1.0113 \\
7  & 0.384650 & 0.063919405 & 0.019634649 & 0.9971 \\
8  & 0.302176 & 0.050171020 & 0.013893049 & 0.9962 \\
9  & 0.242630 & 0.040401389 & 0.010069109 & 0.9991 \\
10 & 0.198636 & 0.033199571 & 0.007480714 & 1.0028 \\
11 & 0.165391 & 0.027740371 & 0.005690717 & 1.0064 \\
12 & 0.139735 & 0.023508123 & 0.004422466 & 1.0094 \\
\end{tabular}
\caption{Fitted dimensionless $\sigma=\sigma_{\rm phys}a^2$,
$\delta$, and the ratio $2N_y\sigma/g_{\rm lat}^2$ along the fixed-$N_y=3$
sequence.}
\label{tab:weak_sigma_delta}
\end{table}

For the seven production points,
$2N_y\sigma/g_{\rm lat}^2=0.9962$--$1.0113$. Thus the fitted coefficient
tracks the thin-strip scale over the production range
$0.139735\leq g_{\rm lat}^2\leq0.5$.

For comparison, the Hamiltonian semiclassical reference is
\begin{equation}
\sigma_{\rm sc}=A\sqrt{g_{\rm lat}^2}\exp[-B/g_{\rm lat}^2],
\end{equation}
where $B=\pi^2v_H(0)=3.172$ and $v_H(0)=0.3214$~\cite{Loan:2002ej}.
We fix $A$ at the coarsest point to compare the coupling dependence in
Fig.~\ref{fig:weak_sigma_scaling}. The data follow the strip scale rather
than the exponential shape and do not isolate a monopole string tension.

At fixed bare coupling, increasing $N_y$ reduces the geometric strip scale
as $1/N_y$, but observing the bulk exponential also requires the transverse
width and charge separations to be sufficiently large, together with
convergence in $\Lambda$.

\subsection{Weak-coupling basis convergence}

At $N_x=4$, $N_y=3$, and $g^2=0.01$, we repeat the vacuum and charged-sector
calculation for $\Lambda=1,\ldots,7$ using $r/a=1,2,3$. The DMRG tolerance
is $10^{-8}$ and the final SVD cutoff is $10^{-10}$. With
$\overline g^2(r/a=2)=4\sigma+2\delta$, Table~\ref{tab:weak_lambda}
shows rapid convergence.
\begin{table}[!htbp]
\centering

\begin{tabular}{ccc}
$\Lambda$ & $\overline g^2(r/a=2)$ &
$|\overline g^2_\Lambda-\overline g^2_7|/\overline g^2_7$ \\
\hline
1 & 0.007796339 & $1.16\times10^{-2}$ \\
2 & 0.007707199 & $4.42\times10^{-5}$ \\
3 & 0.007706860 & $1.56\times10^{-7}$ \\
$4\text{--}7$ & 0.007706858 & $\text{below numerical resolution}$ \\
\end{tabular}
\caption{Weak-coupling local-basis convergence at $g^2=0.01$ on
$4\times3$.}
\label{tab:weak_lambda}
\end{table}

The three separations fix the three fit parameters, leaving no
goodness-of-fit degrees of freedom.

\subsection{Basis dependence of the extrapolation}

We compare the $\Lambda=1$, $N_y=3$ sequence with independently retuned
$\Lambda=2$ sequences at $N_y=3,5$, using the same $u=0.393820$,
$r_1/a=N_x/3$, $r_2=\sqrt{2}r_1$, and potential-fit range
$r/a=1,\ldots,N_x-3$. In addition to $N_x=6,\ldots,12$, we include
$N_x=16,20$ (Table~\ref{tab:continuum_extension}).
The $\Lambda=2$ calculations use DMRG tolerance $10^{-7}$ and final SVD
cutoff $10^{-9}$; the $\Lambda=1$ extension uses $10^{-8}$ and $10^{-10}$.

\begin{table}[H]
\centering

\small
\setlength{\tabcolsep}{3pt}
\begin{tabular}{cccccc}
$\Lambda$ & $N_y$ & $N_x$ & $g^2$ & $\Sigma_s$ & $2N_y\sigma/g^2$ \\
\hline
1 & 3 & 16 & 0.0795385 & 0.7816813 & 1.0174 \\
1 & 3 & 20 & 0.0511681 & 0.7837661 & 1.0216 \\
2 & 3 & 16 & 0.082392 & 0.781168 & 0.9800 \\
2 & 3 & 20 & 0.052894 & 0.783456 & 0.9869 \\
2 & 5 & 16 & 0.133180 & 0.766380 & 0.9438 \\
2 & 5 & 20 & 0.086423 & 0.773399 & 0.9620 \\
\end{tabular}
\caption{Additional $N_x=16,20$ points; the $\Lambda=2$ values are preliminary.}
\label{tab:continuum_extension}\label{tab:lambda2_extension}
\end{table}

Table~\ref{tab:continuum_intercepts_extension} compares the three
extrapolation forms [Eqs.~\eqref{eq:fit3}--\eqref{eq:fit_a2_a4}] before and after adding the
larger lattices. The $N_x\leq12$ fits use $N_x=9,\ldots,12$ for $f_1,f_2$
and $N_x=8,\ldots,12$ for $f_3$; the extended fits use $N_x=11,12,16,20$
and $N_x=10,11,12,16,20$, respectively.

\begin{table}[H]
\centering

\small
\setlength{\tabcolsep}{3pt}
\begin{tabular}{cccccc}
$\Lambda$ & $N_y$ & $N_x^{\max}$ & $f_1(0)$ & $f_2(0)$ & $f_3(0)$ \\
\hline
1 & 3 & 12 & 0.786653 & 0.792559 & 0.788007 \\
1 & 3 & 20 & 0.787364 & 0.789773 & 0.787585 \\
2 & 3 & 12 & 0.787156 & 0.794011 & 0.787804 \\
2 & 3 & 20 & 0.787501 & 0.790192 & 0.787584 \\
2 & 5 & 12 & 0.779217 & 0.795680 & 0.789524 \\
2 & 5 & 20 & 0.784034 & 0.791415 & 0.786314 \\
\end{tabular}
\caption{Step-scaling intercepts before and after adding $N_x=16,20$.}
\label{tab:continuum_intercepts_extension}
\end{table}

At $N_y=3$, increasing $\Lambda$ from 1 to 2 lowers $\Sigma_s$ by
$0.660\%$ at $N_x=6$, $0.133\%$ at $N_x=12$, and $0.0396\%$ at $N_x=20$.
Corresponding intercepts differ by less than $0.2\%$ before the extension
and $0.054\%$ after it (Fig.~\ref{fig:lambda2_continuum}).
At $\Lambda=2$, adding the larger lattices reduces the fit spread from
$0.00685$ to $0.00269$ for $N_y=3$ and from $0.01646$ to $0.00738$ for
$N_y=5$. The largest linear-versus-quadratic interpolation shift in
$\Sigma_s$ is $2.82\times10^{-5}$, not a total uncertainty estimate.

The coarser points are more sensitive to the basis increase than the
extrapolated result. At fixed $u=\sigma R_1^2+\delta R_1$, with $R_1=r_1/a$,
the fitted form gives
\begin{equation}
 \Sigma_s=s^2\sigma R_1^2+s\delta R_1
          =s^2u-s(s-1)\delta R_1.
\end{equation}
If the logarithmic contribution $\delta R_1$ becomes negligible, $\Sigma_s$
approaches $s^2u\simeq0.78764$ independently of the fitted slope $\sigma$.
The similar intercepts are therefore consistent with a linear-potential
regime in both sequences, but do not by themselves show convergence to the
untruncated theory.

}

\end{appendix}

\end{document}